\documentclass[a4paper,11pt]{article}

\usepackage{jheppub}
\usepackage{amssymb}
\usepackage{amsfonts}
\usepackage{amsmath}
\usepackage{mathtools}
\usepackage{bm}
\usepackage{macros}

\title{\boldmath A Type IIB Dualifold}

\author{Zihni Kaan Baykara, Cumrun Vafa}

\affiliation{Jefferson Physical Laboratory, Harvard University,\\
Cambridge, MA 02138, USA}

\emailAdd{zbaykara@g.harvard.edu} \emailAdd{vafa@g.harvard.edu}

\abstract{We present evidence that ten-dimensional type IIB string theory orbifolded by the $\mathbb{Z}_3$ duality symmetry at the strong-coupling fixed point $\tau=\exp(2\pi i/3)$ defines a consistent theory. Motivated by its F-theory dual description and its duality with M-theory in nine dimensions and type IIA string theory in eight dimensions, we argue that the resulting theory is a ten-dimensional chiral non-supersymmetric theory with a discrete gauge symmetry whose bosonic part is the automorphism group of the Narain lattice $\Gamma^{2,2}(A_2)$ ($S_3 \times S_3 \times \mathbb{Z}_2$). We further identify the ten-dimensional tachyon fields with the corresponding Narain moduli. Gauge anomalies cancel nontrivially through a topological Green--Schwarz mechanism involving a $\mathbb{Z}_3$-valued two-form gauge symmetry---the ten-dimensional lift of the familiar zero-form orbifold quantum symmetry in the eight-dimensional type IIA description.
}

\begin{document}

\maketitle
\flushbottom

\section{Introduction}
\label{sec:introduction}
Orbifold construction, where a symmetry in a quantum gravity is eliminated, is a familiar toolkit for constructing new string vacua \cite{Dixon:1985jw,Dixon:1986jc}.  The symmetries are usually taken to be perturbatively visible in string theory.  In principle one can also consider orbifolding by gauge symmetries which emerge only at strong coupling (see e.g. \cite{Kumar:1996zx, Dabholkar:2002sy,Hull:2004in}), though unfortunately there are not many reliable techniques for illuminating the consistency and properties of such orbifolds.  A notable exceptions are supersymmetric F-theory constructions \cite{Vafa:1996xn} which at some points \cite{Dasgupta:1996ij} can be interpreted as orbifolding by non-perturbative dualities of IIB string theory.  In this paper we propose a non-supersymmetric example of such a duality:  Type IIB in 10d orbifolded by the $\mathbb{Z}_3$ duality symmetry at the strong coupling point $\tau=\exp(2\pi i/3)$.

The main tool in this duality is the F-theory picture which suggests its circle compactification to M-theory on $T^2/\mathbb{Z}_3$ in 9 dimensions and its torus compactification to IIA on $T^2/\mathbb{Z}_3$.  We find that the untwisted sector in 10d after the ``dualifold'' by itself would be anomalous.  To find what the `twisted sectors' must provide we follow its relation to IIA compactified on $T^2/\mathbb{Z}_3$.  This leads to the anticipation of a discrete gauge symmetry whose bosonic part is $S_3\times \mathbb{Z}_2$ in 10d, with a matter spectrum which is non-chiral gravitationally but chiral under this gauge symmetry.    In fact we show that the charge lattice of 3-branes is given by the Narain lattice of $\Gamma^{2,2}(A_2)$ which suggests that the 10d gauge symmetry gets further enhanced to a fermionic lift of the isometry group $S_3\times S_3\times \mathbb Z_2$ of the lattice.  Gauge anomalies cancel non-trivially via a topological Green-Schwarz mechanism.  
For type 0B/A we show that the D-brane charges are given by $\Gamma^{1,1}(A_1)$ Narain lattice.  In that case we show that the tachyons of type 0B/A plays the role of Narain moduli of the $\Gamma^{1,1}$ lattice.  Given this we propose that the tachyons of the dualifold model can be naturally identified with the Narain moduli of the $\Gamma^{2,2}$. The spectrum of the theory includes (see Table \ref{tab:iib-fields}), in addition to metric, a pair of self-dual and a pair of anti-self-dual 4-form gauge fields, the tachyons, as well as equal numbers of spinors of each chirality.  There could potentially exist additional neutral scalars.

\begin{table}[h]
\centering
\small
\renewcommand{\arraystretch}{1.2}
\begin{tabular}{cc}
\hline
Field & $S_3\times S_3\times \mathbb Z_2$ rep \\
\hline
$g_{MN}$
& $\mathbf{1}_+$ \\
$D_4^-$
& $(\mathbf1,\mathbf{2})_-$ \\
$D_4^+$
& $(\mathbf{2},\mathbf1)_-$ \\
$\lambda_+$
& $\widehat{\mathbf2}_f\otimes (\mathbf 1,\mathbf 2)_+ $ \\
$\lambda_-$
& $\widehat{\mathbf2}_f\otimes (\mathbf 2,\mathbf 1)_+ $ \\
$T$ & $(\mathbf 2,\mathbf 2)_+ $\\
\hline
\end{tabular}
\caption{The 10d IIB dualifold spectrum in terms of the enhanced discrete gauge symmetry representations. The $S_3\times S_3$ representations are given in parentheses, $\mathbb Z_2$ charge as subscript, and $\widehat{\mathbf 2}_f$ is a representation of the fermionic lift. Chiral fermions are denoted $\lambda_\pm$. Tachyons $T$ can be viewed as the Narain moduli.}
\label{tab:iib-fields}
\end{table}

One may also wonder about other dualifolds in 10d such as quotienting by $\mathbb Z_4,\mathbb Z_6$.  We find evidence that even though they may exist their duality with M-theory and IIA hinted by the F-theory picture does not work in a simple way.

The organization of this paper is as follows:  In section~\ref{sec:F-IIA} we review the F-theory and its connection with M-theory and IIA upon circle and torus compactifications.  In section~\ref{sec:iia-z3}, we study type IIA on $T^2/\mathbb{Z}_3$, including its massless and tachyonic spectrum and its gauge symmetries.  In section~\ref{sec:z3-quotient}, we make our proposal for the spectrum of IIB dualifold under the $\mathbb{Z}_3$ quotient.  In section~\ref{sec:other-dualifolds} we explain why the other potential dualifolds of IIB in 10d are different. We present our conclusions in section~\ref{sec:conclusions}.  Some of the technical details of the anomaly cancellation are discussed in the appendix.

\section{F-theory/M-theory/IIA duality chain}
\label{sec:F-IIA}

The quotient considered in this paper acts on the non-perturbative duality
symmetry of 10d type IIB string theory. It is therefore useful to first pass to the
duality frame in which this action becomes purely geometric.  Indeed F-theory proposes how this works \cite{Vafa:1996xn}:  Compactifying IIB on $S^1$ geometrizes the duality torus identifying it with M-theory compactification on it to 9 dimensions.  Upon a further compactification on a circle this gets related IIA compactified on the duality torus. In the type IIA
description, the type IIB axio-dilaton becomes the complex structure of a
physical torus and the complexified K\"ahler structure of IIA is the complex structure of the IIB compactification torus. So, this IIB/IIA duality is non-perturbative and is not just the usual T-duality, which would have been the exchange of both K\"ahler and complex structures.  Moreover the 10d limit of IIB is obtained by taking area of IIB torus to be large which is mapped to making the coupling of IIA going to infinity. One can then in general consider fibering this picture over some base which leads to the following picture. 
 We consider an elliptic manifold where $T^2$, the duality torus, is fibered over a base $B$, giving a manifold $K=T^2\rtimes B$, then
\begin{align}(\text{F-theory} \ {\rm on} \ K)= (\mathrm{IIB} \ \text{on} \ B )\underset{S^1}{\longrightarrow} ( \text{M-theory}\ {\rm on}\ K)\underset{S^1}{\longrightarrow} (\mathrm{IIA} \ \text{on} \ K)\end{align}

In this paper we would be interested in the case where the fibration is dictated by a quotient of a discrete symmetry.
The dualifolds of type IIB quotient compactified on $T^2$ can then be
studied as an ordinary geometric orbifolds of M-theory and IIA and recovering the 10d IIB theory in the strong coupling limits.

\subsection{Supersymmetric F-theory quotients}
As reviewed above, F-theory brings the duality torus to life.  Indeed some of its supersymmetric quotients have already been studied as the special case of the more general elliptic fibration over the base of F-theory.  

For example in \cite{Sen:1996vd} the F-theory quotient $T^2\times T^2/\mathbb{Z}_2$ where $\mathbb Z_2$ acts on both tori was studied and connected to IIB orientifold on $T^2/\mathbb{Z}_2$, leading to $SO(8)^4$ gauge symmetry.  The supersymmetric quotients $T^2\times T^2/\mathbb Z_{3,4,6}$
were studied in \cite{Dasgupta:1996ij} and shown to lead to $E_6^3,E_7^2\times SO(8), E_8\times E_6\times SO(8)$ respectively.
Other examples studied include quotients of $T^2\times T^2\times T^2/\mathbb Z_n\times \mathbb Z_m$ leading to $(1,0)$ supergravity theories in 6 dimensions \cite{Hayashi:2018iqb}. These in particular not only lead to gauge symmetries, but also conformal matter charged in particular under exceptional groups.
Also local models like $T^2\times \mathbb C^3/\mathbb Z_n$ with D3 brane probes at the singularities were studied in \cite{Garcia-Etxebarria:2015wns,Aharony:2016kai} to realize ${\cal N}=3$ supersymmetric quantum theories in 4d.

All these examples have the following feature:  They identify a particular quotient of the duality torus with action on the visible 10d spacetime.  Moreover the combination is such that some amount of supersymmetry is preserved.  In this paper we would like to generalize this to the case where the action is only on the duality torus, which thus necessarily breaks supersymmetry.  Of course one can also consider other quotients which couple an orbifold action on the duality torus with an action on spacetime which breaks supersymmetry, which we leave to future work.

The closest example of the supersymmetric F-theory quotients that would be similar to what we are studying in this paper is $T^2\times \mathbb C/\mathbb Z_3$ where the quotient acts to rotate the duality torus fiber and the base by 120 degrees.  This leaves a twisted sector in 8d which is localized at the fixed point and has an $E_6$ gauge symmetry.  Here we will be considering the same $\mathbb Z_3$ with no action on $\mathbb C$ i.e. $T^2/\mathbb Z_3 \times \mathbb C$ and ask whether it gives rise to a well defined theory in 10d.

\subsection{Map of parameters}
\label{subsec:F-IIA-fields}

We consider type IIB on $T^2$. Let its complex structure
be $\sigma_B$ and its area in string units $A_B$.
Denote the IIB axio-dilaton 
$\tau_B=C_0+i/g_B$, where $C_0$ is the RR axion and $g_B$ is
the 10d IIB coupling. We set $B_2=C_2=0$ on the torus.\footnote{One way to see these parameter identifications is to denote the IIB torus with $T^2_{xy}$. T-dualizing the $x$ direction gives M-theory on a torus
with coordinates $\tilde x,y,z$, where $\tilde x$ is the T-dual direction and $z$ is the coordinate of the M-theory circle. The torus with coordinates $\tilde x,z$ has complex structure $\tau_B$
\cite{Vafa:1996xn,Schwarz:1995jq}. The area of this torus is inversely
proportional to the IIB $x$ circle radius. We now reduce M-theory on the $y$ circle and get type IIA
on the torus with coordinates
 $\tilde x,z$.  One can also deduce the parameter identification using standard relations in the literature (see e.g. \cite{Kiritsis:1997em}).
Throughout this paper, by type IIA theory we mean the final frame given in the above F-theory duality chain involving IIB theory/M-theory/IIA.}

Let $\sigma_A$ denote the shape modulus of the IIA torus $T^2$ and
$\rho_A=b_A+iv_A$ its complexified K\"ahler modulus,
where $v_A$ is the area of $T^2$, and $b_A$ is
the period of $B_2$. The parameter map is
\begin{equation}
\begin{gathered}
    \sigma_A=\tau_B,\\
    \rho_A=\sigma_B.
    \label{eq:essential-F-IIA-map}
\end{gathered}
\end{equation}
Thus the IIB axio-dilaton determines the IIA torus shape,
while the IIB torus shape determines its K\"ahler class. The 10d IIA coupling $g_A$ is
\begin{equation}
g_A=A_B\sqrt{\operatorname{Im}\tau_B \operatorname{Im}\sigma_B}\ .
    \label{eq:iib-limit-iia-coupling}
\end{equation}

To recover 10d IIB, we take $A_B\to\infty$ at fixed $\tau_B$ and $\sigma_B$. This corresponds to
$g_A\to\infty$ with both IIA torus moduli fixed.
We are interested in the dualifold at the point $\tau_B=\omega$ and we will use its duality to IIA on a hexagonal torus at strong coupling.

\subsection{Symmetries}
\label{subsec:F-IIA-symmetries}

We now study the potential IIA realization of dualities of 10d IIB motivated by the F-theory duality chain. This will help us determine the IIB dualifold as the strong coupling limit of a geometric orbifold of IIA.

The type
IIB S-duality group becomes the large diffeomorphisms of the geometric torus in the type IIA frame. For
\begin{equation}
    \gamma=
    \begin{pmatrix}
        a&b\\ c&d
    \end{pmatrix}
    \in\mathrm{SL}(2,\mathbb Z),
\end{equation}
the action
\begin{equation}
    \tau_B\longmapsto
    \frac{a\tau_B+b}{c\tau_B+d}
\end{equation}
is mapped to the corresponding large diffeomorphism
of the type IIA $T^2$ since $\sigma_A=\tau_B$.

The generator of the $\mathbb Z_3$ dualifold group is
\begin{equation}
    \gamma_3\equiv (ST)^2
    =
    \begin{pmatrix}
        -1&-1\\
         1& 0
    \end{pmatrix},
    \qquad
    \gamma_3^3=1.
    \label{eq:gamma-three}
\end{equation}
It fixes the axiodilaton at
\begin{equation}
    \tau_B=\omega,\qquad 
    \gamma_3\cdot\omega=\omega,
\end{equation}
where $\omega\equiv e^{2\pi i/3}$ is a third root of unity. In the type IIA frame, $\gamma_3$ is simply the
$120^\circ$ rotation of the hexagonal torus.

In addition to the $SL(2,\mathbb Z)$ duality group, there are also two more $\mathbb Z_2$ symmetries of type IIB that enlarge the duality group to $GL(2,\mathbb Z)$. They are generated by orientifold action $\Omega_B$ and left-moving fermion number $(-1)^{F_L}$ with actions as
\begin{equation}
    \Omega_B:
    (B_2,C_2)\longmapsto(-B_2,+C_2),
    \qquad
    (-1)^{F_L}:
    (B_2,C_2)\longmapsto(+B_2,-C_2).
\end{equation}
These symmetries correspond to orienti-reflections in the IIA frame
\begin{equation}
\begin{array}{c|c}
    \text{type IIB} & \text{type IIA} \\ \hline
    \Omega_B
        & \Omega_{A}R_{+} \\[2mm]
    (-1)^{F_L}
        & \Omega_{A}R_-\\[2mm]
    \Omega_B(-1)^{F_L}
        & R_+R_-
\end{array}
\label{tab:parity-duality-map}
\end{equation}
where $R_{+},R_-$ denote reflections of the
corresponding IIA torus directions, and $\Omega_{A},\Omega_B$ denote the orientifold action of IIA and IIB respectively. Note that the exchange of $\Omega_B$ and
$(-1)^{F_L}$ under IIB S-duality is mapped to the exchange
of the two cycles of the IIA torus.\footnote{These are only the perturbative shadows of the actual symmetries: orbifolding by $(-1)^{F_L}$ gives type IIA whereas orbifolding by $\Omega_B$ gives type I, so they are not quite the same.}

On fermions, the $SL(2,\mathbb Z)$ group acts through its metaplectic cover $Mp(2,\mathbb Z)$, which encodes a choice of square root that determines the action on fermions as \cite{Gaberdiel:1998ui, Pantev:2016nze}
\begin{align}
    \psi \mapsto \left(\pm\frac{\sqrt{c\tau+d}}{|\sqrt{c\tau+d}|}\right)^q \psi,\qquad (\gamma,\pm\sqrt{c\tau+d}) \in Mp(2,\mathbb Z),
\end{align}
where gravitino has charge $q=-1$ and dilatino has $q=3$. We are interested in the dualifold of the order three lift of $\gamma_3$
\begin{align}
    \widehat{\gamma}_3 = (\gamma_3, \omega^2),
\end{align}
The other lift of $\gamma_3$ has doubled order and differs from it by an extra multiplicative factor of $(-1)^F$.  

The additional parity elements $\Omega_B,(-1)^{F_L}$ do not commute on fermions
\begin{equation}
    \Omega_B(-1)^{F_L}
    =
    (-1)^F(-1)^{F_L}\Omega_B,
    \label{eq:parity-algebra}
\end{equation}
and so the bosonic $\mathbb Z_2\times \mathbb Z_2$ lifts to fermionic $D_8$. Since the parities square to the identity, the full duality group is the $\mathrm{Pin}^+$ lift
\begin{equation}
    1
    \longrightarrow
    \{1,(-1)^F\}
    \longrightarrow
    \widetilde{\mathrm{GL}}_{\mathrm{Pin}^+}(2,\mathbb Z)
    \longrightarrow
    \mathrm{GL}(2,\mathbb Z)
    \longrightarrow
    1.
    \label{eq:duality-double-cover}
\end{equation}
This is the full duality group of IIB in 10d and equivalently the geometric large diffeomorphism group of IIA on $T^2$.

\section{\texorpdfstring{Type IIA on $T^2/\mathbb Z_3$}
{Type IIA on T2/Z3}}
\label{sec:iia-z3}

Under the IIA/F-theory duality described in section~\ref{sec:F-IIA}, the $\mathbb Z_3$
duality of 10d type IIB becomes a geometric rotation of the
type IIA on $T^2$. We therefore begin with weak coupling type IIA in 8d,
where the spectrum can be determined directly from the orbifold CFT of the worldsheet. We present the spectrum in terms of the representations of the leftover geometric symmetries that act on $T^2/\mathbb Z_3$. We will leverage this remaining discrete gauge symmetry to obtain information about the strong coupling type IIB limit in 10d.

\subsection{Fields}

\label{subsec:iia-z3-fields}

Using a dimensionless complex coordinate $w$, write the
hexagonal torus as
\begin{equation}
    T^2=\mathbb C/\Lambda,
    \qquad
    \Lambda=\mathbb Z+\omega\mathbb Z.
    \label{eq:hexagonal-torus}
\end{equation}
The complex structure is fixed to $\sigma_A=\omega$, and the area remains an independent parameter. Let $J_{89}$
be the rotation generator in the compact $(X^8,X^9)$ plane, normalized so
that $w=X^8+iX^9$ has charge $+1$. The $\mathbb Z_3$ action on the worldsheet is
\begin{equation}
    g=(-1)^{ F}
      \exp\!\left(\frac{2\pi i}{3}J_{89}\right),
    \qquad g^3=1,
    \label{eq:iia-order-three-lift}
\end{equation}
where $ F$ is spacetime fermion number.  Note that the factor $(-1)^{ F}$ is
essential as without it the spin lift has order six rather than three.

We use indices $\mu,\nu=0,\ldots,7$ for the noncompact directions and
$i,j=8,9$ for the torus.  The massless untwisted fields are
\begin{equation}
\begin{aligned}
    &g_{\mu\nu},
    &&\phi,\quad v,\quad b,
    &&A^{(1)}_\mu,\quad A^{(3)}_\mu,
    &&B_{\mu\nu},\quad C_{\mu\nu\rho},
    &&\lambda^a,\quad a=1,2 .
    \label{eq:iia-untwisted-fields}
\end{aligned}
\end{equation}
Here $g_{\mu\nu}$ is the 8d metric, $\phi$ is the
dilaton, $v$ is the area of $T^2$, and
\begin{equation}
    b\equiv\frac{1}{(2\pi\ell_s)^2}\int_{T_A^2}B_2
\end{equation}
is the NSNS axion.  The vector $A^{(1)}_\mu$ descends
from the RR one-form $C_1$, while
\begin{equation}
    A^{(3)}_\mu\equiv C_{\mu 89}
\end{equation}
descends from the RR three-form.  Finally, $B_{\mu\nu}$ and
$C_{\mu\nu\rho}$ are the surviving NSNS two-form and RR three-form,
and the $\lambda^a$ are two 8d non-chiral spin-$\frac12$ fields.  No
gravitino survives, so the orbifold is nonsupersymmetric.

The action has three fixed points
\begin{equation}
    f_0=0,
    \qquad
    f_1=\frac{1+2\omega}{3},
    \qquad
    f_2=\frac{2+\omega}{3}.
    \label{eq:iia-z3-fixed-points}
\end{equation}
At every fixed point $f\in \mathbb Z_3$, the combined $g$- and $g^2$-twisted sectors contain
the following fields:
\begin{equation}
    T_f^{\mathbb C},
    \qquad
    s_f^{\mathbb C},
    \qquad
    A_{\mu,f}^{\mathbb C},
    \qquad
    C_{\mu\nu\rho,f},
    \qquad
    \lambda_f^a\quad(a=1,2).
    \label{eq:iia-twisted-fields}
\end{equation}
The fields $T_f^{\mathbb C}$, $s_f^{\mathbb C}$, and $A_{\mu,f}^{\mathbb C}$ are respectively one
complex tachyon, one complex massless scalar, and one complex vector.  The field
$C_{\mu\nu\rho,f}$ is an unconstrained real three-form, and the $\lambda_f^a$ are two
non-chiral spin-$\frac12$ fields.  The localized tachyon has dimensionless mass
\begin{equation}
    \frac{\alpha'M^2}{4} = - \frac13.
\end{equation}
The remaining fields in \eqref{eq:iia-twisted-fields} are massless.  For the interpretation of the localized tachyon at the fixed point in the non-compact case see \cite{Adams:2001sv} (the compact case is discussed in \cite{vafa2001mirrorsymmetryclosedstring}).\footnote{The mirror worldsheet description is an
$\mathcal N=(2,2)$ Landau--Ginzburg model, with superpotential $W=X^3+Y^3+Z^3+\kappa XYZ+t_1X+t_2Y+t_3Z$, 
where $\kappa$ is the parameter related to the K\"ahler modulus, and $t_1,t_2,t_3$ describe the three complex tachyon deformations.
At $t_i=0$, we have $X,Y,Z$ with weights
$h=\bar h=1/6$, reproducing the tachyon mass $\alpha'M^2/4=h-1/2=-1/3$. In the
$\mathbb C/\mathbb Z_3$ example, the mirror
deformation $W=u^3+tu$ describes smoothing the
orbifold singularity
\cite[sec.~3.1]{vafa2001mirrorsymmetryclosedstring}.} 

\subsection{Discrete symmetries}

\label{subsec:iia-z3-symmetries}
The fractional translation and two orientifold reflections
\begin{equation}
    \mathsf{T}:w\longmapsto w+\frac{1+2\omega}{3},
    \qquad
    \mathsf{P}_\pm=\Omega_A R_\pm,
    \qquad R_\pm:w\longmapsto\pm\overline w
    \label{eq:fractional-translation}
\end{equation}
act on the three fixed points as
\begin{align}
    \mathsf{T}&: f_0\to f_1\to f_2\to f_0,\\
    \mathsf{P}_+&:f_1 \leftrightarrow f_2,\\
    \mathsf{P}_-&:\text{all }f_i\text{ fixed}.
\end{align} Here $\Omega_A$ is worldsheet parity since spacetime parity by itself is not a symmetry of IIA. The orientifold reflections arise as the normalizer of the orbifold group $\mathbb Z_3 \subset \widetilde{GL}_{\mathrm{Pin}^+}(2,\mathbb Z)$, and translations arise as commuting gauge symmetries.

Lifting the symmetries to the worldsheet Hilbert space, they generate the group
\begin{align}
    G_{\mathrm{IIA}}\equiv \langle \mathsf{T},\mathsf{P}_+,\mathsf{P}_-,(-1)^F\rangle \cong \mathbb Z_3 \rtimes D_8,
\end{align}
with bosonic part $S_3\times\mathbb Z_2$.\footnote{$D_8$ denotes the dihedral group of order $8$.} The group relations satisfy
\begin{equation}
\begin{gathered}
    \mathsf T^3=\mathsf{P}_+^2=\mathsf{P}_-^2=1,\qquad (-1)^F\text{ central},\\
    \mathsf{P}_+\mathsf T\mathsf{P}_+^{-1}=\mathsf T^{-1},\qquad
    [\mathsf{P}_-,\mathsf T]=0,\qquad \mathsf{P}_+\mathsf{P}_-=(-1)^F\mathsf{P}_-\mathsf{P}_+.
\end{gathered}
\label{eq:fermionic-reflection-algebra}
\end{equation}
Thus the reflections commute on bosons and anticommute on fermions.

All the irreducible representations of $G_{\mathrm{IIA}}$ are specified below.
Here $\sigma_i$ are Pauli matrices and $I_2$ is the identity. Bold labels give
the complex dimensions of the irreducible representations in the table.
For the bosonic representations, the subscript is the $\mathsf{P}_-$ parity,
and the prime distinguishes the sign representation of $S_3$.
\begin{equation}
\renewcommand{\arraystretch}{1.25}
\begin{array}{c|cccc}
    &\mathsf T&\mathsf{P}_+&\mathsf{P}_-&(-1)^F\\ \hline
    \mathbf1_\pm
        &1&1&\pm 1&1\\
    \mathbf1'_\pm
        &1&-1&\pm 1&1\\
    \mathbf2_\pm
        &\operatorname{diag}(\omega,\omega^2)
        &\sigma_1&\pm  I_2&I_2\\
    \mathbf2_f
        &I_2&\sigma_1&\sigma_3&-I_2\\
    \mathbf2_f^\pm
        &\operatorname{diag}(\omega,\omega^2)
        &\sigma_1&\pm \sigma_3&-I_2
\end{array}
\label{eq:iia-fixed-point-irreps}
\end{equation}
The fermionic doublet $\mathbf2_f$ admits a real structure. The remaining fermionic doublets
$\mathbf2_f^+$ and $\mathbf2_f^-$ are complex conjugates. We denote their real four-dimensional combination
by $\mathbf4_f$, so that its complexification obeys
\begin{align}
    (\mathbf4_f)_{\mathbb C}
    =\mathbf2_f^+\oplus\mathbf2_f^-.
\end{align}
Fourier combinations of the fixed point Hilbert spaces form the modules
\begin{equation}
\begin{aligned}
    \mathbf3_+&=\mathbf1_+\oplus\mathbf2_+,\qquad
    \mathbf3_-&=\mathbf1'_-\oplus\mathbf2_-,\\
    \mathbf6_f&=\mathbf2_f\oplus\mathbf4_f.
\end{aligned}
\label{eq:iia-recurring-fixed-point-modules}
\end{equation}
Here $\mathbf3_\pm$ and $\mathbf6_f$ are reducible real representations
of the indicated dimensions. The module $\mathbf3_+$ is the permutation
representation, and $\mathbf3_-$ is the permutation representation
tensored with the sign. The fermionic module $\mathbf6_f$ is the analog of $\mathbf3_\pm$ for fermions.

The fields of section~\ref{subsec:iia-z3-fields} organize in these representations as
\begin{align}
    \mathcal H_1&={}
    \bigl(g_{\mu\nu}\oplus\phi\oplus v\oplus b
       \oplus A^{(1)}_\mu\oplus A^{(3)}_\mu\bigr)
       \otimes\mathbf1_+\nonumber\\
    &\oplus(B_2\oplus C_3)\otimes\mathbf1'_-\nonumber\\&
       \oplus\lambda\otimes\mathbf2_f,
    \label{eq:iia-untwisted-fixed-point-representations}\\
    \mathcal H_g\oplus\mathcal H_{g^2}&={}
    \bigl( T_{-1/3}^{\mathbb C}\oplus s^{\mathbb C}
       \oplus A_\mu^{\mathbb C}\bigr)\otimes\mathbf3_+
       \nonumber\\
    &\oplus C_3\otimes\mathbf3_-\nonumber\\&
       \oplus\lambda\otimes\mathbf6_f.
    \label{eq:iia-twisted-fixed-point-representations}
\end{align}
Here $\lambda$ is a non-chiral 8d spinor, and the representation
dimensions include all fixed-point multiplicities.
The $g$ and $g^2$ sectors supply CPT-conjugate helicity halves,
combining into the real unconstrained three-forms and non-chiral
fermions. Table~\ref{tab:iia-z3-spectrum} summarizes the spectrum.

\begin{table}[t]
\centering
\small
\renewcommand{\arraystretch}{1.2}
\begin{tabular}{ccc}
\hline
Field & Untwisted & Twisted \\
\hline
$g_{\mu\nu}$
    & $\mathbf1_+$
    & $0$ \\
$\varphi$
    & $3\cdot\mathbf1_+$
    & $2\cdot\mathbf3_+$ \\
$A_\mu$
    & $2\cdot\mathbf1_+$
    & $2\cdot\mathbf3_+$ \\
$B_2$
    & $\mathbf1'_-$
    & $0$ \\
$C_3$
    & $\mathbf1'_-$
    & $\mathbf3_-$ \\
$\lambda$
    & $\mathbf2_f$
    & $\mathbf6_f$ \\
$ T$
    & $0$
    & $2\cdot\mathbf3_+\;$ \\
\hline
\end{tabular}
\caption{Tachyonic and massless spectrum of type IIA on
$T^2/\mathbb Z_3$ at weak coupling. Each nonzero entry gives the
real $G_{\mathrm{IIA}}$ representation. The twisted column combines the $g$- and
$g^2$-twisted sectors at all three fixed points.
The fields $\varphi$ are massless scalars, $C_3$ denotes
unconstrained real three-forms, $\lambda$ denotes non-chiral
8d spinors, and $T$ are tachyons.}
\label{tab:iia-z3-spectrum}
\end{table}

There is also an orbifold quantum $\mathbb Z_3$ symmetry \cite{Vafa:1989ih}, whose generator acts as
$Q\lvert\psi_k\rangle=\omega^k\lvert\psi_k\rangle$ in the
$g^k$-twisted sector, $k=0,1,2$. Since it acts as a scalar in each twisted sector it commutes with
$\mathsf T,\mathsf P_\pm$. In section~\ref{sec:z3-higgsing} we propose that
it lifts to a one-form symmetry in 9d.

\section{Type IIB\texorpdfstring{$/\mathbb{Z}_3$}{/Z3}}
\label{sec:z3-quotient}
\subsection{Spectrum}

We begin with the 10d type IIB fields invariant under the
$\mathbb Z_3$ duality quotient. Using the representations of
$G_{\mathrm{IIA}}\cong\mathbb Z_3\rtimes D_8$ defined in
section~\ref{subsec:iia-z3-symmetries}, the untwisted spectrum is
\begin{equation}
    \mathcal H^{10d}_1
    =
    (g_{MN}\otimes\mathbf1_+)
    \oplus (D_4^-\otimes\mathbf1'_-)
    \oplus (\lambda_+\otimes\mathbf2_f).
    \label{eq:iib-z3-untwisted-spectrum}
\end{equation}
Here $M,N=0,\ldots,9$, and $D_4^\pm$ denotes a real four-form
potential whose five-form field strength obeys
$F_5^\pm=\pm *F_5^\pm$. Our chirality convention assigns $D_4^-$
to the surviving type IIB four-form. The fields $\lambda_\pm$
are 10d Majorana--Weyl spinors; since $\mathbf2_f$
is a two-dimensional real representation,
$\lambda_+\otimes\mathbf2_f$ contains two such spinors of a given chirality.
No gravitino survives.

The untwisted spectrum is gravitationally anomalous by itself.
A consistent 10d theory therefore requires additional
degrees of freedom. We will refer to these collectively as the
``twisted sector,'' although this terminology does not assume a
twisted worldvolume construction of the duality quotient.
Our guide to its field content is the proposed relation to type IIA
on $T^2/\mathbb Z_3$.

First, the untwisted fields already reproduce the corresponding untwisted
8d IIA spectrum. The relevant toroidal reductions are
\begin{align}
\begin{split}
    g_{MN}
    &\longrightarrow
    g_{\mu\nu}\oplus A_\mu^{\mathbb C}\oplus3\varphi,\\
    D_4^\pm
    &\longrightarrow C_3\oplus B_2,\\
    \lambda_+ &\longrightarrow \lambda
    \label{eq:four-form-branching}
\end{split}
\end{align}
where $C_3$ is an unconstrained real
three-form. Note that a chiral 10d Majorana--Weyl spinor $\lambda_\pm$ becomes one
non-chiral 8d spinor $\lambda$, with little-group
branching $\mathbf 8_\pm \to \mathbf 4 \oplus \overline{\mathbf4}$. These fields give precisely IIA untwisted fields in
\eqref{eq:iia-untwisted-fixed-point-representations}.

We next turn to the twisted sector fields suggested by the IIA fixed
points. Their three-forms and fermions transform as
\begin{equation}
   (C_3\otimes\mathbf1'_- )\oplus (C_3\otimes \mathbf 2_-)
    \oplus (\lambda\otimes\mathbf2_f )\oplus (\lambda \otimes \mathbf4_f).
\end{equation}
The three-forms suggest 10d four-form potentials,
and the fermions suggest 10d Majorana--Weyl fields.
Their 8d spectrum does not determine their
10d chiralities, but the discrete representations,
CPT, and gravitational anomaly cancellation uniquely determine the chiral twisted sector as
\begin{align}
    \mathcal H^{10d}_{\mathrm{tw}}
    \supset{}&
    (D_4^-\otimes\mathbf1'_-)
    \oplus (D_4^+\otimes\mathbf2_-)
    \nonumber\\
    &\oplus(\lambda_+\otimes\mathbf2_f)
    \oplus(\lambda_-\otimes
    \mathbf4_f).
    \label{eq:iib-z3-twisted-spectrum}
\end{align}
Table~\ref{tab:iib-z3-spectrum} summarizes the fields and their
multiplicities in the untwisted and twisted sectors.

\begin{table}[t]
\centering
\small
\renewcommand{\arraystretch}{1.2}
\begin{tabular}{ccc}
\hline
Field & Untwisted & Twisted \\
\hline
$g_{MN}$
& $\mathbf{1}_+$
& $0$ \\
$D_4^-$
& $\mathbf{1}'_-$
& $\mathbf{1}'_-$ \\
$D_4^+$
& $0$
& $\mathbf{2}_-$ \\
$\lambda_+$
& $\mathbf{2}_f$
& $\mathbf{2}_f$ \\
$\lambda_-$
& $0$
& $\mathbf{4}_f$ \\
\hline
\end{tabular}
\caption{Proposed 10d IIB dualifold spectrum in terms of $G_{\mathrm{IIA}}$ representations. Scalars are left undetermined.}
\label{tab:iib-z3-spectrum}
\end{table}

There are two four-forms and four Majorana-Weyl fermions of each chirality, so the pure gravitational anomaly cancels directly. The discrete symmetries act chirally, so their gauge anomalies require the detailed analysis given in appendix~\ref{app:dualifold-anomaly} and summarized in section~\ref{sec:anomalies}.

The proposed spectrum reproduces all the IIA three-forms and
fermions. Indeed, their total 8d representations are
$\mathbf1'_-\oplus\mathbf3_-$ and $\mathbf2_f\oplus\mathbf6_f$,
respectively, in both descriptions.
Table~\ref{tab:iib-iia-spectrum-comparison} gives the full comparison
and displays the remaining mismatches.

\begin{table}[htbp]
\centering
\small
\renewcommand{\arraystretch}{1.2}
\begin{tabular}{lcc}
\hline
Field
    & IIA on $T^2/\mathbb Z_3$
    & IIB$/\mathbb Z_3$ on $T^2$ \\
\hline
$g_{\mu\nu}$
    & $\mathbf1_+$
    & $\mathbf1_+$ \\
$\varphi$
    & $3\cdot\mathbf1_+$
    & $3\cdot\mathbf1_+$ \\
$A_\mu$
    & $2\cdot\mathbf1_+$
    & $2\cdot\mathbf1_+$ \\
$C_3$
    & $\mathbf1'_-\oplus\mathbf3_-$
    & $\mathbf1'_-\oplus\mathbf3_-$ \\
$\lambda$
    & $\mathbf2_f\oplus\mathbf6_f$
    & $\mathbf2_f\oplus\mathbf6_f$ \\
\hline
$B_2$
    & $\mathbf1'_-$
    & $\mathbf1'_-\oplus\mathbf3_-$ \\
Localized $A_\mu$
    & $2\cdot \mathbf3_+$
    & $0$ \\
Localized $\varphi$
    & $2\cdot \mathbf3_+$
    & ? \\
Localized $ T$
    & $2\cdot \mathbf3_+$
    & ? \\
\hline
\end{tabular}
\caption{Comparison of the 8d spectra with a delimiter between the matches and the mismatches. Each entry
gives the real $G_{\mathrm{IIA}}$ representation. The spinors $\lambda$ are non-chiral and the three-forms $C_3$ are unconstrained. Undetermined entries denoted by a `$?$' allow possible
additional 10d scalar fields.}
\label{tab:iib-iia-spectrum-comparison}
\end{table}

The localized complex vectors
$A_\mu^{\mathbb C}\otimes\mathbf3_+$ account for the six real
localized vectors in the table. They have no counterpart among
the direct reductions of the 10d fields. We propose that they acquire a
geometric interpretation in the intermediate 9d
M-theory description, where they are absorbed into metric data
localized at the orbifold fixed points. We discuss the motivation
from localized D0-brane charges in
section~\ref{sec:z3-higgsing}.

Conversely, each additional 10d four-form supplies a
two-form as well as the three-form used to construct the lift.
The resulting 8d two-form content is
\begin{equation}
    B_2\otimes
    \bigl(2\cdot\mathbf1'_-\oplus\mathbf2_-\bigr),
    \label{eq:naive-tensor-reduction}
\end{equation}
whereas the perturbative IIA orbifold contains only the NSNS $B_2\otimes\mathbf1'_-$. The surplus is therefore
\begin{equation}
    B_2\otimes
    \bigl(\mathbf1'_-\oplus\mathbf2_-\bigr)
    =
    B_2\otimes\mathbf3_-.
    \label{eq:surplus-8d-tensors}
\end{equation}
These are three additional two-forms. Reconciling the two
descriptions requires them to leave the massless spectrum along the
path to weak type IIA coupling which we discuss in section~\ref{sec:z3-higgsing}.

Finally, the perturbative localized tachyons and massless scalars of IIA
each form a complex copy of $\mathbf3_+=\mathbf1_+\oplus \mathbf2_+$, giving six real fields
in each row of table~\ref{tab:iib-iia-spectrum-comparison}.
Their possible 10d descendants must respect the
surviving discrete symmetry, but their masses in the strong coupling limit are not
fixed by the argument above. In general indeed masses of scalars will be renormalized. However, we still argue for the existence of tachyons in 10d that can account for $2\cdot \mathbf 2_+$ tachyons in section~\ref{sec:enhanced}.

In section~\ref{sec:z3-higgsing} we discuss the proposed resolution of the
gauge-field mismatches and the uncertainty in the scalar sector.
The discrete anomaly calculation will then provide an independent
test of the proposed spectrum.  Before that we now turn to evidence for an enhanced gauge symmetry in the 10d limit.

\subsection{Enhanced gauge symmetry in 10d}
\label{sec:enhanced}
We now provide evidence that the gauge symmetry in 10d is larger than $G_{\mathrm{IIA}}\cong \mathbb Z_3\rtimes D_8$.
We argue it is the fermionic lift of $S_3\times S_3 \times \mathbb Z_2$.
We do this by showing that the 3-brane sector enjoys a bigger symmetry group.\footnote{That the higher dimensional limit can have more symmetries which get broken by holonomies around the circle as we go down to M-theory is a typical phenomenon in F-theory.
For example in the supersymmetric version of the $\mathbb Z_3$ quotient, the higher dimensional theory has an $E_6$ gauge symmetry at each fixed point which gets broken to $SU(3)^3$ in going down on a circle to M-theory.}
\paragraph{3-Brane Charge Lattice:}

We will now show that the 4-dimensional 3-brane charge lattice is given by the Narain lattice 
\begin{equation}
    \Lambda\equiv\Gamma^{2,2}(A_2)
    =\bigl\{(v_L,v_R)\in
    \Lambda_{SU(3)}^\vee\times\Lambda_{SU(3)}^\vee
    \mid v_L-v_R\in\Lambda_{SU(3)}\bigr\},
    \label{eq:appendix-A2-diagonal-gluing}
\end{equation}
where $\Lambda_{SU(3)}$ is the root lattice and $\Lambda_{SU(3)}^\vee$ the weight lattice.  
This is the unique unimodular lattice of signature $(2,2)$ that admits an asymmetric $\mathbb Z_3$ symmetry.
The fact that we have an integral lattice with signature (2,2) which is unimodular is a reflection of Dirac quantization (integrality) for electric/magnetic 3-branes, as well as completeness (unimodularity) and the signature reflects the signature of the 3-branes (self-dual vs anti-self-dual).  To see that it is a unique choice we simply note that the projection to self-dual side should admit an integral lattice with $\mathbb Z_3$ symmetry which fixes it to be an $A_2$ lattice.  The fact that its discriminant is 3, means that the intersection of the lattice with the anti-self-dual states should also have discriminant 3, so that gluing left and right by a $\mathbb Z_3$ identification would lead to a unimodular lattice.  The only 2d lattices with these properties are either the $A_2$ lattice or the rectangular lattice with norms $(1,3)$.  It is easy to check that the latter does not lead to an integral lattice by a $\mathbb Z_3$ gluing with right.  We are thus left with the unique option of the Narain $\Gamma^{2,2}(A_2)$ lattice.  Even though evenness of the lattice was not a requirement it is an outcome of what we have found.  

Moreover we have found that the lattice has a bigger symmetry than expected: $S_3\times S_3\times \mathbb Z_2$, the two Weyl groups and the overall reflection.  This suggests that our theory has a bigger gauge group, namely the spin lift of $S_3\times S_3\times \mathbb Z_2$. We conjecture that the 10d theory enjoys this enhanced gauge symmetry which gets broken when we go down to 9d on a circle due to a gauge holonomy around the circle.

\paragraph{Enhanced symmetry representations: }

We will now identify the fermionic lift and the representations more carefully.

Let us first identify how $G_{\mathrm{IIA}}$ acts on the charge lattice.  Let $c_L$ be a $120^\circ$ rotation of the left $A_2$ lattice and
$s_L$ a Weyl reflection of a simple root. Then
\begin{equation}
\begin{aligned}
    \mathsf T(v_L,v_R)&=(c_Lv_L,v_R),\\
    \mathsf{P}_+(v_L,v_R)&=(-s_L v_L,-v_R),\\
    \mathsf{P}_-(v_L,v_R)&=(-v_L,-v_R).
\end{aligned}
    \label{eq:appendix-narain-group-action}
\end{equation}
Note that $(-1)^F$ acts trivially.

The positive and negative planes carry the representations
\begin{equation}
    \Lambda_{\mathbb R}^{+}\cong
    \mathbf2_-,
    \qquad
    \Lambda_{\mathbb R}^{-}
    =\mathbf1'_-\oplus \mathbf1'_-.
\end{equation}
Thus the positive plane gives a doublet of self-dual fields,
while the negative plane gives two anti-self-dual fields
transforming in the same one-dimensional representation reproducing the four-form spectrum
\begin{align}
    D_4^+\otimes \mathbf2_- \quad \oplus \quad  D_4^-\otimes (\mathbf1'_-\oplus \mathbf1'_-).\label{eq:D4-spectrum-GIIA}
\end{align}
Fermions follow analogous representations except that the lattice reflections exchange two independent internal components
\begin{align}
    \lambda_- \otimes  \mathbf 4_f \quad \oplus \quad \lambda_+\otimes (\mathbf2_f\oplus \mathbf 2_f).\label{eq:fermion-spectrum-GIIA}
\end{align}

Now we consider the enhancement. The full Narain symmetry group includes $c_R,s_R$ actions on the right that are restored in 10d. The bosonic symmetry group is then
\begin{align}
    G_b\equiv S_3 \times S_3 \times \mathbb Z_2 \subset O(2,2,\mathbb Z).
\end{align}
Therefore, the four-form spectrum with respect to the full symmetry group becomes
\begin{align}
    D_4^+\otimes (\mathbf2,\mathbf1)_-\quad \oplus \quad D_4^-\otimes (\mathbf1,\mathbf2)_-,
\end{align}
where in parentheses we denote the $S_3\times S_3$ representation and the subscript is the $\mathbb Z_2$ charge. One can check that decomposition with respect to $S_3\times \mathbb Z_2$ gives \eqref{eq:D4-spectrum-GIIA}.

The fermionic version of the full symmetry group is given by the Pin lift
\begin{align}
    G_f\equiv (\mathbb Z_3 \times \mathbb Z_3)\rtimes (D_8\times \mathbb Z_2) \subset \mathrm{Pin}(2,2,\mathbb Z),\label{eq:pin-lift-Gf}
\end{align}
with reflections $s_L,s_R,(-1)^F$ generating $D_8$ and $s_Ls_R\mathsf{P}_-$ generating a commuting $\mathbb Z_2$ as
\begin{equation}
\begin{gathered}
  s_R^{\,2}=\mathsf{P}_-^{\,2}=1,
 \qquad s_L^{\,2}=(-1)^F,\\
  s_L s_R=(-1)^F s_R s_L,
 \qquad
  s_{L(R)}\mathsf{P}_-=(-1)^F\mathsf{P}_- s_{L(R)},
\end{gathered}
\label{eq:pin22-squares-commutators}
\end{equation}
acting on the $\mathbb Z_3\times \mathbb Z_3$ generated by $c_L,c_R$ as
\begin{equation}
\begin{gathered}
 s_{L(R)} c_{L(R)} s_{L(R)}^{-1}
 = c_{L(R)}^{-1},\qquad [c_{L(R)},s_{R(L)}]=0,\qquad \mathsf{P}_- c_{L(R)}\mathsf{P}_-^{-1}= c_{L(R)},
\end{gathered}
\label{eq:enhanced-fermionic-presentation}
\end{equation}
where each relation holds with all uniformly parenthesized or unparenthesized subscripts. We retain $\mathsf{P}_-$ as the overall inversion $(v_L,v_R)\mapsto (-v_L,-v_R)$. The original generators of the IIA symmetry group can be identified as
\begin{align}
    \mathsf{T}=c_L,\qquad \mathsf{P}_+ = s_L\mathsf{P}_-.
\end{align}
Note that $s_L^2=(-1)^F$ is fixed by $\mathsf{P}_+^2=1$, therefore the Pin lift choice is fixed as $s_R^2=1$.

We propose that the theory has a parity symmetry exchanging chiralities together with exchange of the left and right discrete symmetry groups:\footnote{If parity didn't act as an exchange on the discrete symmetry  groups, the gauge anomaly would vanish trivially. This would have been the case if $D_4^+$ appeared both with $(\mathbf 2,\mathbf 1)_-$ and $(\mathbf 1,\mathbf 2)_-$, which isn't the case.} 
\begin{align}
    D_4^+\leftrightarrow D_4^-,\qquad \lambda_+\leftrightarrow \lambda_-,\qquad S_{3,L}\leftrightarrow S_{3,R},
\end{align}
where subscripts $L,R$ denote the $S_3$ acting on the left or right planes. This fixes the fermionic representations as
\begin{align}
    \lambda_- \otimes \widehat{\mathbf2}_f\otimes (\mathbf 2,\mathbf 1)_+ \quad \oplus \quad \lambda_+\otimes\widehat{\mathbf2}_f\otimes(\mathbf1,\mathbf2)_+,
\end{align}
where the fermionic representation $\widehat{\mathbf 2}_f$ is a lift of $\mathbf 2_f$ with the actions:
\begin{align}
    \begin{array}{c|ccccc}
         &  c_{L,R} & s_L & s_R & \mathsf{P}_- & (-1)^F\\
         \hline
         \widehat{\mathbf2}_f & I_2 & -i\sigma_2 & \sigma_1 & \sigma_3  & -I_2
    \end{array}
\end{align}
In other words, fermions transform under the Weyl group $S_3$ as $\mathbf 2$, with reflections $s_L,s_R$ also exchanging two internal flavors. This extension respecting the proposed parity symmetry and reproducing \eqref{eq:fermion-spectrum-GIIA} is unique.\footnote{There are also other representations that reproduce \eqref{eq:fermion-spectrum-GIIA} but do not respect the parity symmetry.}

\paragraph{Type 0A, 0B and $\Gamma^{1,1}$ D-brane charge lattice:}
It is natural to ask what we know about the charge lattice structure in other non-supersymmetric theories in 10d.  In particular type IIA, IIB orbifolded by $(-1)^F$ lead to the non-supersymmetric type 0A, 0B string theories \cite{Dixon:1986iz}.  These theories have no fermions and have a doubled RR gauge fields compared to their parent type II theory and get an extra real tachyonic field $T$.  Let us denote the RR field (of a given $p$-form) of the original theories by $C$ and the new one coming from the twisted sector by $C'$.  It is known that these theories have additional branes $D^\pm$  charged under the combinations 
\begin{align}C_{\pm}=\frac{1}{\sqrt{2}}(C\pm C'),\end{align}
where the brane from the parent theory is a combination of $D^+D^-$.\footnote{Note that we follow the usual convention that Type 0 fields labeled by $\pm$ do not denote self-dual or anti-self-dual fields, but rather the linear combination used in defining these fields. In fact, the $+$ and $-$ fields exchange under electric-magnetic duality.}

We first want to focus on the type 0B, and in particular on the 4-form gauge fields.  The untwisted 4-form field is anti-self dual $C_4=D_4^-$ whereas the twisted sector is self-dual $C_4'=D_4^+$, consistent with the need for anomaly cancellations.  The analog of the lattice for the 3-brane charges is an integral self-dual $(1,1)$ lattice.  There are only two such lattices: The Narain one $\Gamma^{1,1}$ and the cubic one which is not even, generated by two vectors $(1,0)$ and $(0,1)$ with norms $-1$ and $1$ respectively.  The fact that we have charges in the $(\pm\frac{1}{\sqrt{2}},\pm\frac{1}{\sqrt{2}})$ as generators implies among the two options that the lattice of 3-brane charges is the Narain lattice $\Gamma^{1,1}$.  Moreover this is at the self-dual radius, i.e.
$\Gamma^{1,1}(A_1)$, which enjoys a $\mathbb Z_2\times \mathbb Z_2$ given by the Weyl group of the lattice, as is the symmetry of type 0 theories generated by orientifold $\Omega$ and quantum symmetry $Q$.  Note that in particular the $D3^+$ and $D3^-$ charges correspond to vectors in the $\Gamma^{1,1}(A_1)$ given by $e_1,e_2$ with $e_1^2=e_2^2=0$ and $e_1\cdot e_2=1$.

It is natural to ask if there is a mode in the 0B theory that plays the role of the Narain moduli for $\Gamma^{1,1}(A_1)$.
Indeed the 0B theory has an extra real field, the tachyon $T$. Can that play the role of the Narain moduli? Indeed giving vev to $T$ breaks the $\mathbb Z_2$ quantum symmetry of 0B theory.
On the other hand the $\mathbb Z_2$ quantum symmetry acts on the fields by $C\rightarrow C, C'\rightarrow -C'$ which means that the $D3^+\leftrightarrow D3^-$ which in terms of the Narain lattice is a $\mathbb Z_2$ reflection on the positive norm plane.  In other words giving vev to $T$ should move us away from the $A_1$ symmetric point of $\Gamma^{1,1}$.
Indeed there is already another hint how this works in more detail.  In \cite{Meessen:2001wk} it was shown that if one uses the two combined $C_\pm$ which is neither self-dual nor anti-self dual, one can write an effective action for the 5-forms field strength of the form
\begin{align}\int f^+(T) F_+^5\wedge  * F_+^5+ f^-(T) F_-^5\wedge *F_-^5\end{align}
where
\begin{align}f^+(T)=1/f^-(T).\end{align}
$f^+(T)$ has been computed up to second order in $T$ and in a convenient normalization of $T$ (which differs from \cite{Garousi:1999fu,Baykara:2026gem,Baykara:2026vdc} by a factor of $\sqrt{2}$) we find 
\begin{align}f^\pm(T)=1\pm2T+2T^2+...\end{align}
Similar to the conjecture made by Garousi \cite{Garousi:1999fu}, we conjecture that this sums up to
\begin{align}f^\pm(T)=\frac{1\pm T}{1\mp T}.\end{align}
This means that the tachyon dependence of the kinetic term can be absorbed by redefining
\begin{align}F^5_\pm \rightarrow \sqrt{\frac{1\pm T}{1\mp T}}F_\pm^5. \end{align}
Viewing $F^5_\pm $ as the two null directions of the $\Gamma^{1,1}$ lattice, indeed this transformation can be viewed as a boost!  Because in 2d the boost by a velocity $v$ in light cone parameterization $ds^2=dx^+dx^-$ is given by the transformation
\begin{align}dx^\pm \rightarrow \gamma (1\pm v)dx^{\pm}=\sqrt{\frac{1\pm v}{1\mp v}}dx^{\pm}. \end{align}
So identifying $T=v$ we indeed have an $SO(1,1)$ Lorentz boost of the $\Gamma^{1,1}$ lattice.  Indeed this structure also works for other gauge forms of both 0B and 0A where the corresponding charge lattices all are given by the $\Gamma^{1,1}$ Narain lattice.  It is interesting that in this context the bound proposed in \cite{Baykara:2026gem} $|T|<1$ gets related to $|v|<1$, which is the statement that the boost speed should be less than the speed of light giving further evidence for that conjecture.\footnote{This identification may suggest an $SO(1,1)$ invariant metic for T kinetic term given by $\frac{1}{(1-T^2)^2}\nabla T\nabla T$.
In the 0A setup of \cite{Baykara:2026gem} if we denote the radii of the two circles by $R_{1,2}$ then from M-theory picture at large radii the canonical metric will have a term $\sum_i dR_i^2/R_i^2$, which by changing variables to $R^2=R_1R_2$ and $T=(R_1-R_2)/(R_1+R_2)$ leads to this kinetic term for $T$.  However this form of kinetic term is motivated at large radii and it is natural to expect corrections at small radii corresponding to weak string coupling, which should happen if $T=1$ is to be at finite distance expected in \cite{Baykara:2026gem}.  This is also consistent with the absence of correction for kinetic term for four point function of tachyon at genus 0 \cite{Garousi:2003db}.}

\paragraph{Tachyons of $\mathrm{IIB}/\mathbb Z_3$ and Narain moduli of $\Gamma^{2,2}$:}

Having seen the role of tachyons of type 0 theories as Narain moduli of the D-brane charge lattice, it is natural to conjecture that for the dualifold $\mathrm{IIB}/\mathbb Z_3$ we must also have tachyonic fields which move the Narain moduli for $\Gamma^{2,2}$.  In that case we would need the tachyonic field to parameterize $SO(2,2)/SO(2)\times SO(2)$ which means that we would expect a $(\mathbf 2,\mathbf 2)_+$ tachyon field, which is naturally parameterized by two complex fields $T_1,T_2$ each parameterizing a unit disk:  $|T_1|\leq 1, |T_2|\leq 1$.
This will account at least for the IIA tachyons which transform as 2 copies of the $\bf 2_+$ of $S_3\times \mathbb Z_2$.

\paragraph{Enhanced symmetry spectrum:}
\begin{table}[t]
\centering
\small
\renewcommand{\arraystretch}{1.2}
\begin{tabular}{cc}
\hline
Field & $S_3\times S_3\times \mathbb Z_2$ rep \\
\hline
$g_{MN}$
& $\mathbf{1}_+$ \\
$D_4^-$
& $(\mathbf1,\mathbf{2})_-$ \\
$D_4^+$
& $(\mathbf{2},\mathbf1)_-$ \\
$\lambda_+$
& $\widehat{\mathbf2}_f\otimes (\mathbf 1,\mathbf 2)_+ $ \\
$\lambda_-$
& $\widehat{\mathbf2}_f\otimes (\mathbf 2,\mathbf 1)_+ $ \\
$T$ & $(\mathbf 2,\mathbf 2)_+ \oplus \;?$\\
$\varphi$ & $?$\\
\hline
\end{tabular}
\caption{The 10d IIB dualifold spectrum in terms of the enhanced discrete gauge symmetry representations. The $S_3\times S_3$ representations are given in parentheses, $\mathbb Z_2$ charge as subscript, and $\widehat{\mathbf 2}_f$ is a representation of the fermionic lift. Scalars $\varphi$ are undetermined. Tachyons $T$ are undetermined except for the Narain moduli.}
\label{tab:iib-spectrum-enh}
\end{table}

In light of the discussions of the enhanced gauge symmetry, we finalize the spectrum proposal in terms of the fermionic group $G_f$ in Table~\ref{tab:iib-spectrum-enh}. Note that the restored symmetries in 10d mix untwisted and twisted fields and so the spectrum is given without such a split. This is not unprecedented: the type 0A orientifold symmetry $\Omega$ also exchanges the untwisted $p$-form fields $C$ with twisted $p$-form fields $C'$.

\subsection{The resolution of spectrum mismatch}
\label{sec:z3-higgsing}
So far we have focused on the chiral content of the spectrum in 10d captured by 4-form gauge fields and fermions and matched it to the 8d content on the corresponding IIA theory predicted by F-theory.   Here we would like to discuss the fate of the rest.

We have the untwisted sector that already matches the invariant IIB fields from the untwisted sector. The other fields we need to discuss (see Table \ref{tab:iib-iia-spectrum-comparison}) are the appearance of extra gauge fields in the IIA side and the scalar fields (extra tachyons and massless fields of the twisted sector).  On the other hand from the IIB side if we compactify on $T^2$ in addition to the 3-form gauge fields we have matched on the IIA side, we get 3 additional $B$-fields, one for each of the twisted 4-form fields, where we set two of the directions of 4-form fields to be along the $T^2$, which are absent on the IIA side.

As for the scalar fields we can certainly add them in 10d theory as part of the field content.  We do not know their mass, so in particular they could have become massive or (stay or become) tachyonic. But this would not be any issue in the F-theory duality prediction.  We have already argued that it is natural to expect in 10d to have a $(\mathbf 2,\mathbf 2)_+$ tachyonic field based on the Narain moduli of the charge lattice which could account for 4 of the tachyonic fields on the IIA side.

 The more serious issue to address is the gauge field mismatch.  Let us first discuss the 2-form mismatch coming from the three extra $B$-fields \eqref{eq:surplus-8d-tensors} coming from compactification of the twisted 4-forms in 10d.  To gain an understanding of this, we should recall the intermediate F-theory duality prediction that IIB dualifold compactified on a circle should be equivalent to M-theory on $T^2/\mathbb Z_3$.  Indeed we would expect to get the localized 3-form gauge fields at the fixed points of the $\mathbb Z_3$ action.  We see that the duality between M-theory in 9d and its circle compactification to IIA in 8d also predicts extra 2-form fields localized at the fixed points coming from taking one of the components of the 3-form fields to be along the circle.  In 9d the M2 branes which are charged with respect to the fixed point $C$-fields must come from M2 branes pinned at those points.  Indeed if we take 1 or 2 M2 branes at the fixed points before quotient, the quotient would lead to fractional M2 branes which can be viewed as charged under the two extra $C$-fields, as would be expected from the IIA perspective of the corresponding fractional D2 branes in Douglas-Moore construction \cite{Douglas:1996sw}.
So to get strings charged under the extra $B$-fields localized at the fixed points, we need to wrap these localized M2 branes on the circle.  Let us consider the tension of the resulting string as a function of the radius $R$ which takes M-theory to IIA.  In this case at least for large $R$ in M-theory frame we expect the tension of the wrapped M2 brane to go like
\begin{align}T\sim a R -\frac{b}{R^2}\end{align}
where the first term is the usual tension of stretched M2 brane and the second term is the Casimir contribution coming from the fact that on the wrapped M2 brane the supersymmetry is broken and we have more bosons (for some computable $O(1)$ numbers $a,b$).  From this formula we may expect that at some $R\sim O(1)$ in 11d Planck units the string may become tensionless.  Of course we cannot fully trust the computation at the Planckian length, but at least it is plausible that at some radius it may become tensionless.  If this is the case, then it would not be surprising that the $B$-fields may be Higgsed by the condensation of the tensionless string \`a la \cite{Rey:1989ti,Seiberg:1996vs,Ganor:1996mu}, which was also important in the proposal of \cite[sec.~4.1.3]{Baykara:2026vdc} for establishing Bergman-Gaberdiel duality \cite{Bergman:1997rf}.  

So the only remaining mismatches are the localized gauge fields of IIA for the twisted sectors.  
Before discussing these, let us recall the fate of the extra IIA $U(1)$ gauge field when we go to 11 dimensions.  In the limit of infinite radius the coupling of the gauge field becomes infinitely weak and we absorb it to the 11d metric (the usual KK reduction of metric leading to gauge field).
With this in mind let us ask what are the charged states with respect to extra gauge fields?  They would be 1 or 2 D0 branes stuck at the fixed points.  So it is natural to expect that the fate of the corresponding gauge fields is the same as the usual D0 branes of IIA and that as we go to 9 dimensions the extra gauge fields be absorbed as part of the metric data of the 9d M-theory, and in particular of the metric modes localized at the fixed points.  

We can also revisit the extra IIA orbifold quantum $\mathbb Z_3$ symmetry \cite{Vafa:1989ih}.
Perturbatively, it acts with opposite phases on $C_3^+$ and $C_3^-$. This appears difficult to reconcile with an
ordinary internal symmetry in 9d, where
these polarizations belong to the same field $D_4^\pm$. However,
the compactification has singled out a direction.
A one-form $\mathbb Z_3$ gauge symmetry in 9d can retain this information:
its component along the M-theory circle becomes an
ordinary zero-form $\mathbb Z_3$ transformation after reduction.

It thus seems that the duality can lead to a reasonable match of spectrum for IIB for both M-theory and  IIA compactified on  $T^2/\mathbb Z_3$ as anticipated from F-theory picture.

\subsection{Anomalies and topological Green-Schwarz}
\label{sec:anomalies}
The enhanced gauge symmetry $G_f$ which is the fermionic lift of $S_{3,L}\times S_{3,R} \times \mathbb Z_2$ acts chirally,\footnote{We denote the factor of $S_3$ that acts on the left (right) plane of $\Gamma^{2,2}$ with a subscript $L (R)$.} therefore the corresponding gauge anomalies should be checked.

The anomaly of a 10d theory is calculated by taking a one parameter deformations of it and bringing it back up to a gauge transformation. This amplitude is computed by an 11d anomaly theory, which gives a $\mathbb R/\mathbb Z$ character $\alpha$ on the bordism group $\Omega_{11}^H$
of manifolds equipped with tangential structure
\begin{align}
    H \equiv \frac{\mathrm{Spin}\times G_f}{ \langle (-1,(-1)^F)\rangle}.
\end{align}
In appendix~\ref{app:dualifold-anomaly} we find that the bordism group on the odd-order generators is
\begin{align}
    (\Omega_{11}^H)_{(3)}\cong (\mathbb Z_{27})^2 \oplus \mathbb Z_9 \oplus (\mathbb Z_3)^2,
\end{align}
where each $S_{3,L(R)}$ has anomaly valued in $\mathbb Z_{27}\oplus \mathbb Z_3$, and $\mathbb Z_9$ measures the mixed anomaly between the $S_{3,L}$ and $S_{3,R}$ factors. We find that the anomaly character is
\begin{align}
    \alpha_{(3)} = \left(-\frac13,\frac13,0,-\frac13,\frac13\right).
\end{align}
Therefore each $S_{3,L(R)}$ has an order $3$ anomaly, and there is no mixed anomaly.

For the even order anomalies we consider anomalies of the subgroup $D_8\times \mathbb Z_2$. Note that fermions are non-chiral under these symmetries and so they do not contribute to the anomaly, as shown in appendix~\ref{app:two-primary}. However, the tensor anomaly is notoriously difficult to evaluate for even order elements. Even in the usual type IIB case it is not fully known \cite{Debray:2021vob}. Nonetheless, in appendix~\ref{app:two-primary} we show that under a simplification there exists a choice for which the tensor anomaly vanishes. 

The anomaly character can be written on the 11d manifold $Y$ as
\begin{align}
    \alpha(Y)=\frac13\int_Y X_4\smile Y_7,
\end{align}
where $X_4$ and $Y_7$ are characteristic classes given in appendix~\ref{app:T-GS}. To cancel this anomaly, we employ topological Green-Schwarz mechanism \cite{Garcia-Etxebarria:2017crf,Kobayashi:2019lep}: we need a 10d 3-form $\mathbb Z_3$ gauge field $c_3$ that couples to the $G_f$ background as
\begin{align}
    -\frac13\int_X c_3\smile Y_7
\end{align}
and satisfies a discrete Bianchi identity
\begin{align}
    \delta c_3 = X_4.\label{eq:discrete-bianchi}
\end{align}

Indeed, we have exactly such a candidate gauge field already in the dualifold. Recall in section~\ref{sec:z3-higgsing} we argued that the 0-form  orbifold quantum $\mathbb Z_3$ symmetry $Q$ in 8d IIA lifts to a 1-form symmetry in 9d M-theory. By a similar reasoning, in the F-theory limit, the quantum symmetry lifts to a 2-form symmetry in 10d. Therefore the corresponding 3-form gauge field $c_3$ is exactly the right candidate to cancel the anomalies via the topological Green-Schwarz mechanism.

Note that the discrete Bianchi identity \eqref{eq:discrete-bianchi} forbids some backgrounds one might naively consider acceptable. In particular, a manifold with nonzero mod-three Pontryagin class $\bar p_1$ is not allowed unless an $S_3$-bundle is turned on that cancels it. This is analogous to nonzero $\mathrm{tr}R^2$ requiring a compensating non-trivial gauge bundle in heterotic theory.

\subsection{Non-supersymmetric 4d CFT's}
It is natural to ask what lives on the D3 branes in this dualifold.  There are four 3-branes which we have proposed given by a choice of a lattice vector in $\Gamma^{2,2}(A_2)$. The original D3 brane charge would correspond to  a root vector $R$ in the anti-self-dual part of the lattice, with $R^2=-2$. If we consider $N$ coincident D3 branes this would correspond to $NR$.  Before we quotient we have $SU(N)$ Yang-Mills living on the branes.  To quotient, we have frozen the type IIB and the YM coupling at the strong coupling point $\tau=\exp(2\pi i/3)$ and the self-duality of ${\cal N}=4$ SYM allows us to quotient it.  This is analogous to what was done in \cite{Garcia-Etxebarria:2015wns,Aharony:2016kai} to obtain ${\cal N}=3$ supersymmetric quantum field theories when it was combined with a $\mathbb{Z}_3$ subgroup of the $SO(6)$ R-symmetry group which acted on the 6d space transverse to the D3 brane. The novelty here is that we are not acting on the space at all, and continue to enjoy $SO(6)$ symmetry.  But supersymmetry is broken as the gluinos do transform non-trivially under the quotient and are all projected out.  The orbifold action identifies the electric and magnetic degrees of freedom by a $\mathbb{Z}_3$ rotation.  So we expect to obtain a strongly coupled non-supersymmetric theory in 4 dimensions with $SO(6)$ global symmetry.  It is natural to expect this theory to be conformal.  At least there is no obvious coupling which flows because the gauge coupling $\tau$ is frozen.

It would be interesting to flesh out properties of this strongly coupled non-supersymmetric theory in 4d that we predict to exist.  We do not expect a Lagrangian description for it, given that it is at a strongly coupled point of ${\cal N}=4$ SYM.  It would also be interesting more generally to find what lives on a 3-brane corresponding to an arbitrary charge given by a vector in $\Gamma^{2,2}(A_2)$.

\subsection{dS or AdS?}
\label{subsec:iib-z3-potential}

It would be natural to wonder whether the 10d dualifold has positive or negative vacuum energy.  To get a hint about this we evaluate the
one-loop vacuum amplitude of IIA on $T^2/\mathbb Z_3$. Since the perturbative
spectrum contains twisted-sector tachyons, the modular integral
is divergent. We define it using the string-theoretic $i\varepsilon$
prescription employed by \cite{Baccianti:2025gll}.

The dependence on the type IIA K\"ahler modulus can be determined before
performing the integral. Denote the worldsheet modular parameter by
$\zeta=x+iy$, and write the orbifold sum as
\begin{equation}
    \mathcal Z_{\mathrm{orb}}(\zeta)
    =\frac13\sum_{h,k=0}^{2}
    Z_{h,k}(\zeta)=\sum_{\substack{h,k=0,1,2\\(h,k)\ne(0,0)}}
    \frac{
      \left|
        \vartheta\!\begin{bmatrix}
          \frac12+\frac h3\\[1mm]
          \frac12+\frac k3
        \end{bmatrix}(\zeta)
      \right|^6}
      {|\eta(\zeta)|^{18}},
    \label{eq:potential-orbifold-sum}
\end{equation}
where $h$ labels the spatial twist and $k$ the temporal insertion.
Only $Z_{0,0}$ contains the unrestricted momentum-winding lattice sum.
This term is the vacuum trace of the supersymmetric parent theory
type IIA on $T^2$, and vanishes after the spin-structure sum.
For $k=1,2$, the lattice trace in $Z_{0,k}$ receives contributions only
from charges fixed by $g^k$. The rotation has no nonzero invariant
momentum or winding vector, so this trace reduces to the zero charge
contribution. The remaining sectors are modular images of these
inserted traces and likewise contain no K\"ahler dependent lattice sum.
Consequently, $\mathcal Z_{\mathrm{orb}}$ is independent of $\rho_A$.

The 8d string-frame vacuum energy density is
\begin{equation}
    \mathcal V
    =-\frac{1}{2(2\pi\ell_s)^8}
      \int_{\mathcal F_{i\varepsilon}}
      \frac{d^2\zeta}{y^5}\,\mathcal Z_{\mathrm{orb}}(\zeta).
    \label{eq:potential-regulated-integral}
\end{equation}
We obtain numerically
\begin{equation}
    \ell_s^8\mathcal V
    \approx -1.309585\times10^{-5}
            -i\,4.977089\times10^{-5}.
    \label{eq:potential-one-loop-value}
\end{equation}
The real part is therefore negative and independent of the K\"ahler class.\footnote{In Einstein frame, there will be a scaling dependence on the volume, but the sign remains the same.} The imaginary part records the
instability associated with the six real localized tachyons and agrees
with their masses $\alpha'M^2=-4/3$.\footnote{A tachyon with $m^2=-\mu^2<0$ contributes
$\frac12\int \frac{d^8p}{(2\pi)^8}\log(p^2-\mu^2-i0)$
to the one-loop potential, where $p$ is Euclidean momentum.
For $p^2<\mu^2$, the logarithm has imaginary part $-\pi$.
The six real localized tachyons, each with
$\mu^2=4/(3\alpha')$, therefore give
$
    \operatorname{Im}\mathcal V
    =-\frac{6\pi}{2}
      \int_{p^2<4/(3\alpha')}
      \frac{d^8p}{(2\pi)^8}
    =-\frac{1}{648\pi^3(\alpha')^4}.
$
}

The 10d IIB limit corresponds
to $g_A\to\infty$ at fixed type IIA torus moduli, with
$\sigma_A=\tau_B$ and $\rho_A=\sigma_B$. The negative one-loop sign is
therefore the same for every finite fixed choice of the IIB
compactification-torus shape. This suggests that the
10d dualifold has a negative vacuum energy and therefore an $\mathrm{AdS}_{10}$ critical point, although in principle it is possible that higher loop corrections can change the sign at strong coupling.

\subsection{Tachyon geometry}
\label{subsec:tachyon-condensation}

We now study the moduli space of tachyons identified with Narain parameters as in
section~\ref{sec:enhanced}. The identification suggests that tachyons transform under the bosonic Narain duality group
\begin{align}
    O(2,2,\mathbb Z) \cong (SL(2,\mathbb Z)\times_{\mathbb Z_2}SL(2,\mathbb Z))\rtimes \mathbb Z_2^2,
\end{align}
where the dualifold $(T_1,T_2)=(\omega,\omega)$ is the fixed point of the subgroup $S_3\times S_3 \times \mathbb Z_2$.

Note that the tachyon moduli space does not necessarily have the usual Narain metric of $\frac{O(2,2)}{O(2)\times O(2)}$.  This is similar to the case of 0B for which one may have naively expected the standard coset metric of $\frac{O(1,1)}{O(1)\times O(1)}$, but the tachyon which parametrizes this space is expected to lead to a finite length \cite{Baykara:2026gem} unlike the infinite length the group quotient would have suggested. Similarly in the present case we may expect a finite volume metric for this space unlike what is suggested by the quotient.  For example, if we assume we have
a flat metric, modular invariance of the tachyon metric (which is the symmetry of the theory) can be achieved by taking
\begin{equation}
    ds^2=
    |\eta(T_1)|^8\,dT_1\,d\overline{T}_1+|\eta(T_2)|^8\,dT_2\,d\overline{T}_2.
    \label{eq:tachyon-flat-metric}
\end{equation}
In fact, this metric is unique up to
normalization under the assumptions of flatness, smoothness
and nondegeneracy on $\mathbb H$, and finite volume.

With the metric in \eqref{eq:tachyon-flat-metric}, each tachyon moduli space corresponds to $ T^2/\mathbb{Z}_6$ with the points $T_{1,2}=\omega,i,i\infty$ corresponding to $\mathbb Z_3,\mathbb Z_4,\mathbb Z_6$ fixed points.\footnote{On the scalar manifold, the $\mathbb Z_4$ acts as a $\mathbb Z_2$ since $S^2=-I$ acts trivially on $T_{1,2}$.} To see this, choose the flat coordinate 
\begin{align}
    z(\tau)=x+e^{\pi i/3}y = C\int_{i\infty}^\tau \eta(t)^4 dt
\end{align}
where $C$ is a normalization so that $x\sim x+1,y\sim y+1$ and $2z(i)=1$. Then we can see $T$ is a $2\pi/6$ rotation around the origin, $S$ is a $\pi$ rotation around $z(i)$, and $ST$ is a $2\pi/3$ rotation around $z(\omega)$
\begin{align}
    z(T\tau) =e^{\pi i/3} z(\tau),\quad z(S\tau) -z(i)= -[z(\tau)-z(i)], \quad z(ST\tau)-z(\omega)=\omega^2[z(\tau)-z(\omega)].
\end{align}
It also follows that $ST^3$ and $TST^2$ form the translations for the hexagonal lattice
\begin{align}
    z(ST^3\tau)-z(\tau)=2z(i),\qquad z(TST^2\tau)-z(\tau)=2 z(i)e^{\pi i/3}.
\end{align}
Therefore the tachyons $(T_1,T_2)$ take values in $T^2/\mathbb Z_6\times T^2/\mathbb Z_6$ and there are at least 9 critical points corresponding to $T_{1,2}=\omega,i,\infty$.

It is also natural to ask whether the fermions can acquire
masses along such deformations. One can check that the representations in
Table~\ref{tab:iib-spectrum-enh} allow local Yukawa couplings between the tachyon and the fermions. Globally, one needs the charge of fermions $\lambda_\pm$ under $SL(2,\mathbb Z)\times SL(2,\mathbb Z)$. For example, a charge assignment consistent with the finite group is $(\pm 1,2)$. Using Eisenstein functions, one can construct fermion mass matrices consistent with these charges.

These examples suggest studying whether tachyon condensation in different limits
can lead to type 0B or supersymmetric IIB, as also motivated
by the type 0 discussion in section~\ref{sec:enhanced} and
\cite{Baykara:2026gem}. In particular, the
$\mathbb C/\mathbb Z_6$ cusp of both tachyons suggests that it can be IIB
at $\tau_B=\omega$, whose $SL(2,\mathbb Z)$ stabilizer has
order six. We leave the fate of tachyon condensation as an open problem.

\section{Other Type IIB Dualifolds in Ten Dimensions}
\label{sec:other-dualifolds}

So far we have focused on the quotient of type IIB string theory by the
order-three duality symmetry and on its proposed relation, after circle and
torus compactification, to M-theory and type IIA string theory on
$T^2/\mathbb Z_3$.  It is natural to ask whether the same construction
extends to the other finite duality subgroups
\begin{equation}
    \mathbb Z_2,\qquad \mathbb Z_4,\qquad \mathbb Z_6.
    \label{eq:other-finite-duality-subgroups}
\end{equation}
There is an immediate difference from the order-three case.  When lifted to
the full string Hilbert space, each of these even-order actions doubles because it contains
$(-1)^F$.  Equivalently, the corresponding action on fermions has twice the
order of its bosonic image.  One may likewise double the order-three
quotient by adjoining $(-1)^F$.

Before examining the individual cases, we emphasize that a duality between
two theories need not survive quotients that act only on their internal
degrees of freedom.  For example, quotienting type IIA by $(-1)^{F_L}$
produces type IIB, but the resulting theory does not inherit the naive
11d lift suggested by the type IIA/M-theory duality
\cite{Baykara:2026gem}.  This differs from the familiar use of an
adiabatic argument in fiberwise duality: there the duality is fibered over a
common spacetime on which the quotient acts geometrically
\cite{Vafa:1995gm}.  In the present setting the quotient acts instead
on the internal duality degrees of freedom.  The agreement found for the
$\mathbb Z_3$ dualifold is therefore not guaranteed and should be viewed as a nontrivial result, rather than an
automatic consequence of the parent F-theory duality.

\subsection{The $\mathbb Z_2$ quotient}
\label{subsec:z2-dualifold}

Consider first the central order-two element of the type IIB
duality group.  It leaves the axio-dilaton arbitrary and acts on the
two-form potentials as
\begin{equation}
    (B_{NS},B_R)\longmapsto(-B_{NS},-B_{R}).
    \label{eq:z2-action-two-forms}
\end{equation}
Its lift $\widehat\gamma_2$ to the full type IIB theory has order four,
since
\begin{equation}
    \widehat\gamma_2^{\,2}=(-1)^F.
    \label{eq:z2-lift-square}
\end{equation}
Because this transformation does not fix the coupling, it can be studied
at weak coupling, and indeed it admits such a description \cite{Bossard:2024mls}: a perturbative representative is
$\Omega_B(-1)^{F_L}$, where $\Omega_B$ is the orientifold, and whose action on the two-form potentials agrees with
\eqref{eq:z2-action-two-forms} and squares to $(-1)^F$.  The quotient
may also be viewed as an order-two quotient of type 0B.  Its light spectrum
is that of type 0B with the two-form potentials projected out; in
particular, it retains a non-chiral pair of four-form gauge fields,
\begin{equation}
    D_4^-\oplus D_4^+.
    \label{eq:z2-four-form-spectrum}
\end{equation}
The resulting 10d theory is therefore non-chiral and free of
local gravitational anomalies.

If the geometric F-theory argument commuted with this quotient, compactifying
on $T^2$ would instead lead to type IIA on $T^2/\mathbb Z_2$.  The latter
contains 6 three-form gauge fields.  These cannot be obtained by reducing
the four-form content in \eqref{eq:z2-four-form-spectrum}.  The mismatch is
evidence against a direct extension of the proposed F-theory duality to the
$\mathbb Z_2$ dualifold.  If such a duality exists it must be more subtle.

\subsection{The $\mathbb Z_4$ and $\mathbb Z_6$ quotients}
\label{subsec:z4-z6-dualifolds}

For the $\mathbb Z_4$ and $\mathbb Z_6$ dualifolds, we do not presently know
whether the 10d quotients exist as consistent theories.
Nevertheless, one can assume their existence and test whether the same
geometric duality could hold.

For the $\mathbb Z_4$ quotient, type IIA on $T^2/\mathbb Z_4$ contains 9
three-form gauge fields.  If all of them lifted to 10d
four-form gauge fields, their odd multiplicity would necessarily give a
chiral four-form spectrum.  On the other hand, the quotient projects out
all spacetime fermions because $(-1)^F$ belongs to the orbifold group.
There are consequently no fermionic contributions available to cancel the
local gravitational anomaly of the chiral four-forms.  This provides a
second obstruction to the naive F-theory lift.

The $\mathbb Z_6$ orbifold contains an even number of three-form gauge
fields, so this parity argument alone does not rule out a non-chiral
10d lift.  We nevertheless have no positive evidence for such
a lift, and the failures of the $\mathbb Z_2$ and $\mathbb Z_4$ cases show
that it cannot be inferred from the parent duality.  Finally, doubling the
$\mathbb Z_3$ quotient by adjoining $(-1)^F$ pairs the proposed
$\mathbb Z_3$ bosonic spectrum with a copy of opposite chirality.

These conclusions assume that the three-form gauge fields visible in eight
dimensions survive to strong coupling and lift to four-form gauge fields in
ten dimensions.  In principle, one might imagine that they instead become
massive somewhere along the strong-coupling trajectory.  Unlike the
tensionless string-Higgs mechanism discussed in section~\ref{sec:z3-higgsing}, however,
we know of no corresponding mechanism in the present setup that removes
these three-form fields.  Subject to this assumption, the spectrum and
anomaly arguments above provide evidence that the simple geometric duality
special to the $\mathbb Z_3$ dualifold does not extend to the even-order
quotients.

\label{sec:analysis}

\section{Conclusions}
\label{sec:conclusions}
In this paper we have extended F-theory duality to a non-supersymmetric quotient of type IIB given by a duality action.
We have not only found evidence for the duality to F-theory and IIA upon compactifications to 9 and 8 dimensions, but also believe the fact that the IIA theory makes sense on $T^2/\mathbb{Z}_3$ leads to evidence that this dualifold in 10d is consistent.  Non-trivial evidence involves the cancellation of gravitational and gauge anomalies, where the latter gauge symmetry involves a chiral representation of the group $S_3\times \mathbb Z_2$ (or its extension with an extra $S_3$).  Furthermore we have explained how the mismatch between the light fields of IIA vs. the compactification of IIB dualifold can in principle be resolved.

This work makes it natural to expect more non-supersymmetric dualifolds to exist.  To find exactly what are the consistency conditions for them to exist and what properties they have would be very interesting, not only in its own right, but also for possible applications to the construction of the non-supersymmetric landscape we live in!

\section*{Acknowledgments}
We thank M. Garousi, V. Nevoa and S. Raman for valuable discussions.
This work is supported in part by a grant from the Simons Foundation (602883, CV) and the DellaPietra Foundation.

The main ideas in this paper were developed independently of the LLMs though we have benefited from utilizing the AI systems (Claude and ChatGPT) for routine tasks and some of the calculations in the paper.  We have checked all the calculations for accuracy.

\appendix

\providecommand{\Z}{\mathbb Z}
\providecommand{\R}{\mathbb R}
\providecommand{\C}{\mathbb C}
\providecommand{\HP}{\mathbb H\mathrm P}
\providecommand{\CP}{\mathbb C\mathrm P}
\providecommand{\Spin}{\operatorname{Spin}}
\providecommand{\Pin}{\operatorname{Pin}}
\providecommand{\Tor}{\operatorname{Tor}}
\providecommand{\Arf}{\operatorname{Arf}}
\providecommand{\eF}{\widehat E}
\providecommand{\etal}{\overline\eta_{\mathrm D}}
\providecommand{\etas}{\eta_{\mathrm{sig}}}
\providecommand{\cQ}{\mathfrak c}

\section{Anomaly calculation}
\label{app:dualifold-anomaly}

In this appendix we calculate the anomaly of the enhanced discrete gauge symmetry of the IIB/$\Z_3$ dualifold theory.  

For the odd order bordism generators, the anomaly is fixed using a $\mathbb Z_2$ symmetry of the theory that inverts all fluxes. However, the same argument doesn't work for the even order bordism generators and we will need to assume an explicit construction.

\subsection{Result}
\label{app:dualifold-statement}

The anomaly of a 10d chiral theory is calculated by an 11d invertible theory by taking an auxiliary direction and applying gauge transformations. Allowing topology changes as well, our task is to evaluate the anomaly theory on each bordism class of 11d manifolds
\begin{align}
    \alpha([Y])\in \mathbb R/\mathbb Z,\qquad [Y] \in \Omega_{11}^H
\end{align}
equipped with the tangential and discrete gauge bundle structure
\begin{align}
    H \equiv \frac{\mathrm{Spin}\times G_f}{ \langle(-1,(-1)^F)\rangle},
\end{align}
where $G_f$ is the fermionic enhanced gauge group given in \eqref{eq:pin-lift-Gf}, and so one needs to identify $(-1)^F\in G_f$ with $-1\in \mathrm{Spin}(11)$. 

There are three contributions to the anomaly: Majorana--Weyl
fermions $\lambda_\pm$, self-dual four-form fields $D_4^\pm$, and the sum over
flux sectors.  The last term is subtle because a general
duality covariant quantization of fluxes is not known. For example, type IIB fluxes are quantized in K-theory, which is not compatible with S-duality.

We argue that this microscopic uncertainty does
not affect the three-primary anomaly by a symmetry argument that inverts all the fluxes $x\mapsto -x$. On the three-primary bordism group
\begin{equation}
 (\Omega_{11}^{H})_{(3)}
 \cong (\Z_{27})^2\oplus\Z_9\oplus(\Z_3)^2,
 \label{app:dual:eq:bordism-answer}
\end{equation}
we find the anomaly character
\begin{equation}
 \alpha_{(3)}
 =\left(- \frac13, \frac13,0,-\frac13,\frac13\right).
 \label{app:dual:eq:character-answer}
\end{equation}
In particular, each $S_3$ factor contributes a $\Z_{27}\oplus \Z_3$, whereas $\Z_9$ detects the mixed anomaly.

For the even order bordism generators, we need an explicit flux quantization scheme. For simplicity, we assume fluxes are quantized in ordinary cohomology and find that the anomaly character vanishes
\begin{align}
    \alpha_{(2)}=0.
\end{align}
There is no anomaly at primes larger
than three, so this analysis gives the complete anomaly character.

\subsection{Conventions}

Let $\etal$ be the reduced Dirac eta invariant, and let $\etas$ be the unreduced eta
invariant of the odd-signature operator $\widetilde{B}_{\mathrm{sig}}=-*d$. Majorana--Weyl fermions contribute a factor $\frac 1 2 \etal$,
whereas the self-dual Gaussian carries $-\frac 1 8 \etas$, see
\cite{Hsieh:2020jpj, Witten:2019bou}.  The full anomaly that we compute is
\begin{align}
 \alpha(Y,P)
 &=\frac12\left[
  \etal\bigl(Y;\widehat{\mathbf2}_f\otimes(\mathbf2,\mathbf1)_+\bigr)
 -\etal\bigl(Y;\widehat{\mathbf2}_f\otimes(\mathbf1,\mathbf2)_+\bigr)\right]
 \notag\\
 &\quad-\frac18\left[
  \etas\bigl(Y;(\mathbf1,\mathbf2)_-\bigr)
 -\etas\bigl(Y;(\mathbf2,\mathbf1)_-\bigr)\right]
 +\Arf(q_{Y,P})\pmod1.
 \label{app:dual:eq:anomaly-functional}
\end{align}
Here, the second argument of the eta invariant corresponds to the representation of the field, and $P$ is the $G_b$-bundle. We denote the first two analytic contributions by $\alpha_\eta$.

If $\ell,r,z\in H^1(BG_b;\Z_2)$ denote the reflection classes for $s_L,s_R,\mathsf{P}_-$, its
central extension is
\begin{equation}
 e=(\ell+r)(\ell+z)\in H^2(BG_b;\Z_2).
 \label{app:dual:eq:extension-class}
\end{equation}
An allowed eleven-dimensional background has tangential group
\begin{equation}
 H=\frac{\Spin\times G_f}{\langle(-1,f)\rangle},
 \qquad w_2(TY)+e(P)=0.
 \label{app:dual:eq:tangential-group}
\end{equation}
Thus a failure of $TY$ to be spin may be cancelled by the failure of
the discrete $G_b$-bundle $P$ to lift to $G_f$.

\subsection{Arf on odd-order bordisms}

For a candidate flux quantization scheme $\cQ$, let the discrete fluxes form a
finite abelian group $T_\cQ(Y,P)$.  We assume that the normalized action
$q_\cQ:T_\cQ\to\R/\Z$ satisfies $q_\cQ(0)=0$ and has nondegenerate
polarization
\begin{equation}
 b_\cQ(x,y)=q_\cQ(x+y)-q_\cQ(x)-q_\cQ(y).
\end{equation}
The finite factor is
\begin{equation}
 \frac1{\sqrt{|T_\cQ|}}
 \sum_{x\in T_\cQ}e^{2\pi i q_\cQ(x)}
 =e^{2\pi i\Arf(q_\cQ)}.
 \label{app:dual:eq:gauss-sum}
\end{equation}

Note that we identified $\mathsf{P}_-$ as an inversion on the three-brane charge lattice
\begin{align}
    \mathsf{P}_-(v_L,v_R)= (-v_L,-v_R).
\end{align}
It is reasonable that this symmetry, or some extension of it, acts as inversion on the full flux group. Since $\mathsf{P}_-$ commutes with $c_L,c_R$, this then applies on every background $P$ with holonomy in
$\Z_{3,L}\times \Z_{3,R}$. In particular, this implies that the normalized action is invariant for $P$ backgrounds
\begin{equation}
 q_\cQ(-x)=q_\cQ(x).
 \label{app:dual:eq:inversion}
\end{equation}
In particular, although arbitrary quadratic refinements usually don't satisfy this, their failure to do so arises from a topological coupling of the form
\begin{align}
    \int F_5 \smile c
\end{align}
for some $c$ background. However, $\mathsf{P}_-$ inverts $F_5\mapsto -F_5$ but commutes with the $\mathbb Z_3$ backgrounds we consider and so $c\mapsto c$. This means such a term would not be $\mathsf{P}_-$ invariant and therefore quadratic form $q$ satisfies \eqref{app:dual:eq:inversion} for $\mathsf{P}_-$ to remain a symmetry.

Polarizing $x+(-x)=0$ gives
$2q_\cQ(x)=b_\cQ(x,x)$ and hence
\begin{equation}
 q_\cQ(mx)=m^2q_\cQ(x),\qquad m\in\Z.
 \label{app:dual:eq:homogeneous}
\end{equation}
Thus $q_\cQ$ is homogeneous.  Then the finite Gauss--Milgram theorem \cite{MilnorHusemoller} implies
that its normalized Gauss sum is an eighth root of unity
\begin{equation}
 8\Arf(q_\cQ)\equiv 0 \pmod{1}.
 \label{app:dual:eq:eighth-root}
\end{equation}

Suppose now that $[Y,P]$ has odd bordism order $n$.  Bordism invariance of the complete phase gives
$n(\alpha_\eta+\Arf(q_\cQ))=0$, while \eqref{app:dual:eq:eighth-root} gives $8\Arf(q_\cQ)=0$.
If $un\equiv1\pmod8$, then
\begin{equation}
 \Arf(q_\cQ)=-un\alpha_\eta\pmod1.
 \label{app:dual:eq:rigidity}
\end{equation}
This fixes the Arf contribution on
the selected representatives. However this argument only works for odd order bordism generators, so we will need to choose an explicit quantization to evaluate it at even order generators.

\subsection{Three-primary anomaly}
\label{app:dual:sec:three}

At the prime three, we can ignore the $(-1)^F$ and the $\mathsf{P}_-$.  The $H$-bordism group therefore
reduces to oriented bordism with $S_3\times S_3$ bundles
\begin{equation}
 (\Omega_{11}^{H})_{(3)}
 \cong\Omega_{11}^{SO}
   (BS_{3,L}\times BS_{3,R})_{(3)}.
 \label{app:dual:eq:odd-reduction}
\end{equation}
We now split the bordism group between the two factors
\begin{align}
    \Omega_{11}^{SO}(BS_3\times BS_3) \cong \Omega_{11}^{SO}(\mathrm{pt}) \oplus \widetilde{\Omega}_{11}^{SO}(BS_3)\oplus \widetilde{\Omega}_{11}^{SO}(BS_3)\oplus \widetilde{\Omega}_{11}^{SO}(BS_3\wedge BS_3),
\end{align}
where $\widetilde\Omega$ reduced bordism.

Using \cite[Theorem 4.5]{Shibata1974Metacyclic}, we directly find that the three-primary part of the oriented bordism group for $S_3$ bundles is
\begin{align}
    \widetilde{\Omega}_{11}^{SO}(BS_3)_{(3)} \cong \mathbb Z_{27}\{L_3^{11}\}\oplus \mathbb Z_3\{\HP^2\times L_3^3\},
\end{align}
where brackets denote the generators and $L_3^{2d-1}=S^{2d-1}/\mathbb Z_3$ is the lens space.\footnote{Note that the reference uses $\CP^4$, but it is interchangeable with $\HP^2$ since $[\HP^2]=3[\CP^2\times \CP^2]-2[\CP^4]$ in $(\Omega_8^{SO})_{(3)}$. Multiplying by $[L_3^3]$ then kills the first term and we get $[\HP^2\times L_3^3]=[\CP^4\times L_3^3]$.}

We now separately calculate the mixed anomaly detector $\widetilde{\Omega}_{11}^{SO}(BS_3\wedge BS_3)$. Using the three-local Brown--Peterson splitting \cite[Eq. (12.9)]{Debray:2023yrs} after subtracting the point, we have
\begin{align}
    \widetilde{\Omega}_{11}^{SO}(BS_3\wedge BS_3)_{(3)} \cong \widetilde{BP}_{11}(BS_3\wedge BS_3) \oplus \widetilde{BP}_{3}(BS_3 \wedge BS_3).\label{eq:BP-split}
\end{align}
Now using \cite[Prop 3.13 and Prop 5.4]{JOHNSON1973327} we get a K\"unneth exact sequence
\begin{align}
    (\widetilde{BP}_*(BS_3)\otimes_{BP_*} \widetilde{BP}_*(BS_3))_{11}\to\widetilde{BP}_{11}(BS_3 \wedge BS_3)\to \mathrm{Tor}_1^{BP_*}(\widetilde{BP}_*(BS_3),\widetilde{BP}_*(BS_3))_{10}.
\end{align}
We know what the left factor is by \cite[Theorem 14.3]{Debray:2023yrs}:
\begin{align}
    \widetilde{BP}_k(BS_3)\cong \begin{cases}
    \mathbb Z_3 & k=3\\
    \mathbb Z_9 & k=7\\
    \mathbb Z_{27} & k=11\\
    0, & \text{all other }k\leq 11.
    \end{cases}
\end{align}
Therefore the left factor has no degree 11 and vanishes. We then compute the $\mathrm{Tor}$ group and get
\begin{align}
    \widetilde{BP}_{11}(BS_3\wedge BS_3) \cong \mathrm{Tor}_1^{BP_*}(\widetilde{BP}_*(BS_3),\widetilde{BP}_*(BS_3))_{10}\cong \mathbb Z_9.
\end{align}
For the $\widetilde{BP}_3$ term in \eqref{eq:BP-split}, the same K\"unneth exact sequence applies, but both left and right factors vanish. Therefore we get
\begin{align}
    \widetilde{\Omega}_{11}^{SO}(BS_3\wedge BS_3)_{(3)} \cong \mathbb Z_9.
\end{align}
The generator is given by the lens space $L_3^{11}$ with the diagonal $\Z_3\times \Z_3$ bundle turned on, subtracted by the lens space $L_3^{11}$ with each $\Z_3 \subset S_3$ turned on separately:
\begin{align}
    M_9=[L_\Delta]-[L_L]-[L_R],
\end{align}
where $L_{L(R)}$ denote $L_3^{11}$ with $\Z_{3,L(R)}$ bundle and $L_\Delta$ is the diagonal $\Z_3$ bundle. This concludes the bordism computation and generators of
\begin{align}
    (\Omega_{11}^H)_{(3)} \cong  (\Z_{27})^2\oplus \mathbb Z_9 \oplus (\Z_3)^2.
\end{align}

Now we evaluate the three-primary anomaly. We decompose the representations with respect to the $\mathbb Z_3$ holonomy of each bordism generator. Denote their charges by $q=0,1,2$. Then the eta invariants are given by \cite[Appendix D]{Debray:2021vob}
\begin{align}
 \overline\eta^D_q(L_3^{2d-1})
 &=-\frac1{3i^d}\sum_{j=1}^2
 \frac{e^{-2\pi iqj/3}}
      {(2\sin(\pi j/3))^d},\notag\\
 \eta_{\mathrm{sig},q}(L_3^{2d-1})
 &=-\frac1{3i^d}\sum_{j=1}^2
 \frac{e^{-2\pi iqj/3}}
      {\tan^d(\pi j/3)}.
 \label{app:dual:eq:lens-etas}
\end{align}  

On $L=L_3^{11}$, both invariants equal $2/81$ for
$q=0$ and $-1/81$ for $q=1,2$.  On
$Q=\HP^2\times L_3^3$, the Dirac term vanishes because
$\operatorname{index}(D_{\HP^2})=0$, whereas $\operatorname{Sign}(\HP^2)=1$. Combining these values for the dualifold spectrum gives
\begin{equation}
\begin{array}{c|@{\quad}c@{\quad}c@{\quad}c@{\quad}c@{\quad}c}
 Y&\text{order}&\etal/2&-\eta_{\mathrm{sig}}/8&\Arf(q)&\alpha\\\hline
 L_L&27&-2/27&-1/108&-1/4&-1/3\\
 L_R&27& 2/27& 1/108& 1/4& 1/3\\
 M_9& 9& 0&0&0&0\\
 Q_L& 3& 0&-1/12&-1/4&-1/3\\
 Q_R& 3& 0& 1/12& 1/4& 1/3\\
\end{array}
\label{app:dual:eq:anomaly-table}
\end{equation}

\subsection{Two-primary anomaly}
\label{app:two-primary}
For two-primary anomalies, we can ignore $c_L,c_R$ and only consider
\begin{align}
    D_8\times \mathbb Z_2 \cong \langle s_L,s_R,\mathsf{P}_-,(-1)^F\rangle.
\end{align}
Under these generators, the fermion representations are
\begin{equation}
\begin{array}{c|ccc}
 &s_L&s_R&\mathsf P_-\\ \hline
\widehat{\mathbf2}_f\otimes(\mathbf2,\mathbf1)_+
&
(-i\sigma_2)\oplus(i\sigma_2)
&
\sigma_1\oplus\sigma_1
&
\sigma_3\oplus\sigma_3
\\[2pt]
\widehat{\mathbf2}_f\otimes(\mathbf1,\mathbf2)_+
&
(-i\sigma_2)\oplus(-i\sigma_2)
&
\sigma_1\oplus(-\sigma_1)
&
\sigma_3\oplus\sigma_3
\end{array}.
\end{equation}
One can check that conjugation by $I_2\oplus\sigma_3$ takes the first row to the second. This shows that as $D_8\times \mathbb Z_2$ representations they are isomorphic. Therefore the fermion spectrum is non-chiral and anomaly vanishes trivially.

Therefore it remains to compute
\begin{align}
    -\frac18 \eta_{\mathrm{sig}}(Y) +\mathrm{Arf}(q_Y).
\end{align}
In general, the quadratic form is unknown even for the supersymmetric type IIB case \cite{Debray:2021vob}. If we make the naive simplification that the field is quantized in ordinary cohomology then we can choose a quadratic form refining the torsion linking pairing
\begin{align}
    q_Y=q_{\ell,z}\oplus(-q_{r,z}),
\qquad
q_{\rho,z}([C])
=-\frac{B\cdot\tau_\rho C}{2n}\pmod{ 1},
\qquad
\partial B=nC,
\end{align}
where $C$ is a torsion 5-cycle on the double cover 
\begin{align}
    p_\rho:M\to Y
\end{align}
given by the $s_L,s_R$ holonomies $\rho=\ell,r$, with coefficients in $p_\rho^*\mathbb Z_z$ where $z$ is the holonomy of $\mathsf{P}_-$, and $\tau_\rho$ is the deck transformation. One can check that for this choice the Arf contribution cancels the eta invariant for any $Y$ with a $D_8\times \mathbb Z_2$ background.

\section{Topological Green-Schwarz mechanism}
\label{app:T-GS}
A nonzero anomaly for a gauge symmetry is not necessarily a problem as Green-Schwarz mechanism can cancel it. For type IIB, it was found in \cite{Debray:2021vob} that type IIB has an anomaly for the discrete duality symmetry $\mathbb Z_3\subset SL(2,\mathbb Z)$. 

Several different Green-Schwarz mechanisms for discrete symmetries were proposed to cancel the $\mathbb Z_3$ anomaly of IIB. A natural resolution proposed was a coupling of the $\mathbb Z_3$ duality background to the self-dual field
\begin{align}
    \int F_5 \smile Y_5(a),
\end{align}
where $Y_5(a)$ depends on the $\mathbb Z_3$ duality bundle $a$ turned on. For the dualifold at hand, this Green-Schwarz mechanism is not possible, as we have a symmetry $\mathsf{P}_-$ that commutes with our $\mathbb Z_3$ (whereas it does not with the $\mathbb Z_3$ of type IIB). Such an interaction would not respect the $\mathsf{P}_-$ symmetry. Therefore, we turn to other methods proposed for IIB anomalies. This also motivates the assumption that the quadratic form is homogeneous for the $\mathbb Z_3$ backgrounds.

We find that we can most naturally cancel anomalies using topological Green-Schwarz mechanism \cite{Garcia-Etxebarria:2017crf,Kobayashi:2019lep} as follows. The dualifold anomaly character can be written as
\begin{align}
    \alpha(Y) = \frac13 \int_Y X_4 \smile Y_7,\label{eq:char-class}
\end{align}
where
\begin{align}
    X_4 &\equiv \bar p_1 +u_L+u_R,\\
    Y_7 &\equiv X_4\smile(\gamma_R-\gamma_L),
\end{align}
with $\bar p_1=p_1\pmod{3}$ as the mod-3 Pontryagin class and the classes
\begin{align}
    u_{L(R)} \in H^4(BS_{3,L(R)},\mathbb Z_3),\qquad \gamma_{L(R)}\in H^3(BS_{3,L(R)},\mathbb Z_3),
\end{align}
whose restrictions to $B\mathbb Z_{3,L(R)}$ are
\begin{align}
    &t_{L(R)}=\beta_{\mathbb Z}(a_{L(R)}),\qquad a_{L(R)}\in H^1(B\Z_{3,L(R)},\mathbb Z_3),\\
&\gamma_{L(R)}=a_{L(R)}\smile\overline t_{L(R)},
\qquad u_{L(R)}=\overline t_{L(R)}^{\,2}.
\end{align}
Here the bar denotes reduction modulo three, $\beta_{\mathbb Z}$ is the Bockstein, and $a_{L(R)}$ is the generator. We fix the normalization
\begin{align}
\int_{L_3^3}\gamma_{L(R)}=1,
\qquad
\int_{L_3^{11}}\gamma_{L(R)} u_{L(R)}^2=1.
\end{align}
One can check that evaluating \eqref{eq:char-class} on the bordism generators reproduces \eqref{app:dual:eq:character-answer}.

To cancel the anomaly, we need a term in the 10d action as
\begin{align}
    -\frac13 \int c_3\smile Y_7,
\end{align}
where $c_3$ is a 3-form gauge field with its discrete Bianchi identity 
\begin{align}
    \delta c_3 = X_4.
\end{align}
This is a discrete analogue of the heterotic GS relation $dH= \mathrm{tr}F^2 -\mathrm{tr}R^2$. Indeed as we have noted we do have such a 3-form gauge field implementing the 2-form quantum symmetry in IIB, which comes from the 1-form quantum symmetry in 9d M-theory.

\bibliographystyle{JHEP}
\bibliography{references}

\end{document}